\documentclass[aps,pre,superscriptaddress,amsmath,amssymb,amsfonts,twocolumn,showpacs]{revtex4-2}
\usepackage[utf8]{inputenc}
\usepackage{amsmath}
\usepackage{amssymb}
\usepackage{amsfonts}
\usepackage{amstext}
\usepackage{amsthm}
\usepackage{mathtools}
\usepackage{graphicx}
\usepackage{color}
\usepackage[colorlinks=true,allcolors=blue]{hyperref}
\usepackage{natbib}
\usepackage{xcolor}
\usepackage{float}
\usepackage{mathtools}
\usepackage{physics}
\def\bse{\begin{subequations}}
\def\ese{\end{subequations}}
\def\be{\begin{equation}}
\def\ee{\end{equation}}
\let\~=\tilde
\makeatletter
\let\@=\mathbf
\makeatother
\usepackage{fourier,times}

\begin{document}
\def\thetitle{Nonlinear Stage of Modulational Instability in Repulsive Two-Component\\Bose-Einstein Condensates}

\title{\thetitle}
\author{S. Mossman}
\affiliation{Department of Physics and Biophysics, University of San Diego, San Diego, CA 92110}
\affiliation{Department of Physics and Astronomy, Washington State University, Pullman, Washington 99164-2814}

\author{S. I. Mistakidis}
\affiliation{Department of Physics and LAMOR, Missouri University of Science and Technology, Rolla, MO 65409, USA}

\author{G. C. Katsimiga}
\affiliation{Department of Physics and LAMOR, Missouri University of Science and Technology, Rolla, MO 65409, USA}

\author{A. Romero-Ros}
\affiliation{Departament de Física Quàntica i Astrofísica (FQA), Universitat de Barcelona (UB),  c. Martí i Franqués, 1, 08028 Barcelona, Spain}
\affiliation{Institut de Ciències del Cosmos (ICCUB), Universitat de Barcelona (UB), c. Martí i Franqués, 1, 08028 Barcelona, Spain}

\author{G. Biondini}
\affiliation{Department of Mathematics, State University of New York, Buffalo, New York 14260, USA}

\author{P. Schmelcher}
\affiliation{Center for Optical Quantum Technologies, Department of Physics,
    University of Hamburg, Luruper Chaussee 149, 22761 Hamburg,	Germany}
\affiliation{The Hamburg Centre for Ultrafast Imaging,
    University of Hamburg, Luruper Chaussee 149, 22761 Hamburg,	Germany}

\author{P. Engels}
\affiliation{Department of Physics and Astronomy, Washington State University, Pullman, Washington 99164-2814}

\author{P. G. Kevrekidis}
\affiliation{Department of Mathematics and Statistics, University of Massachusetts Amherst, Amherst, MA 01003-4515, USA}

\date{\today}

\begin{abstract}
Modulational instability (MI) is a fundamental phenomenon in the study of nonlinear dynamics, spanning diverse areas such as shallow water waves, optics, and ultracold atomic gases. 
In particular, the nonlinear stage of MI has recently been a topic of intense exploration,
and has been shown to manifest, in many cases, in the generation of dispersive shock waves (DSWs).
In this work, we experimentally probe the MI dynamics in an immiscible two-component ultracold atomic gas with exclusively repulsive interactions, catalyzed by a hard-wall-like boundary produced by a repulsive optical barrier.
We analytically describe the expansion rate of the DSWs in this system, generalized to arbitrary intercomponent interaction strengths and species ratios.
We observe excellent agreement among the analytical results, an effective 1D numerical model, full 3D numerical simulations, and experimental data.
Additionally, we extend this scenario to the interaction between two counterpropagating DSWs, which leads to the production of Peregrine soliton structures.
These results further demonstrate the versatility of atomic platforms toward the controlled realization of DSWs and rogue waves.
\end{abstract}

\maketitle

\paragraph*{Introduction.}

One of the most universal characteristics of nonlinear media, whether in water waves, acoustics, or nonlinear optics, is the susceptibility of plane waves to minor disturbances, known as modulational instability (MI)~\cite{PhysicaD238p540}.
MI is a topic of active theoretical and experimental investigation, both in its own right~\cite{PhysRevA.96.041601} and as a tool to produce solitary waves across dimensions~\cite{PhysRevLett.125.250401} and fields of application~\cite{Kevrekidis2015,Kivshar2003,ablowitz2,kono}. 

The linear stage of MI is characterized by an exponential growth of perturbations with small wave numbers, as verified by linearizing the governing equations around a uniform background. 
However, when such perturbations grow to a size comparable to the background, their dynamics enter a more complex regime referred to as the nonlinear stage of MI~\cite{ZakhPRL,GinoDion}.
The simplest dynamical model for the description of MI is the single-component self-focusing nonlinear Schr\"odinger (NLS) equation in one dimension with cubic nonlinearity.
It was conjectured in Ref.~\cite{EL1993357} that the nonlinear stage of MI in the NLS equation is described by certain self-similar solutions of the corresponding Whitham modulation equations~\cite{whitham1965non}. 
Such solutions were shown to arise from a large class of initial conditions~\cite{GinoDion} using the inverse scattering transform for the focusing NLS equation with nonzero background~\cite{JMP55p031506}. 
The nonlinear stage of MI was later observed in optical fiber experiments~\cite{Kraych2019, COPIE2020100037}, water waves~\cite{Suret1}, and similar phenomena were, then, also shown to arise in more general NLS-type systems~\cite{SIREV2018v60p888}. 

Ultracold atom quantum simulators provide highly controllable platforms that enable the realization of complex nonequilibrium phenomena, thanks to their exquisite tunability in terms of system parameters~\cite{Bloch_many_body,gross2017quantum,Mistakidis_review}. 
In this context, MI has been examined predominantly in the setting of single-component gases, as a means of generating bright soliton trains~\cite{Strecker2002,Strecker2003}.
Real-time imaging has further elucidated these soliton features (growing sidebands, adjacent soliton phase etc.) in more recent studies~\cite{nguyen,everitt}.
In higher dimensions, MI was also leveraged to induce fragmentation, producing Townes solitons in two dimensions~\cite{Lung_Townes2D,Lung_Townes} and self-patterned Townes-soliton necklaces in the presence of a vortex~\cite{banerjee2024collapse}.
However, the nonlinear stage of MI has not yet been observed in repulsive (defocusing) multicomponent atomic systems.
The latter setting has recently been exploited for a diverse palette of experiments to study effectively attractive dynamics, ranging from the formation of Townes solitons~\cite{Kartashov_review} in quasi-two-dimensional condensates~\cite{Bakkali-Hassani2021} to that of rogue waves in the form of Peregrine solitons (PS) in effectively one-dimensional systems~\cite{our2024PRL}.
The interplay of the repulsive interactions and the resulting effective dynamics of the full system, as well as the richness of the multicomponent setting, render this a particularly useful and versatile platform.

In this Letter, we experimentally and theoretically demonstrate the emergence of MI in a defocusing multicomponent medium consisting of two hyperfine states of a $^{87}$Rb Bose-Einstein condensate (BEC).
Specifically, we analytically characterize the nonlinear stage of MI in this system by investigating the dynamical effects produced by the presence of a hard repulsive barrier enforcing a dam-break problem~\cite{whitham2011linear,EL201611}, namely step-like initial conditions with zero velocity.
Previous studies (starting with the theoretical analysis
of Ref.~\cite{Dutton2005}) have primarily focused on asymmetric settings~\cite{Bakkali-Hassani2021,our2024PRL,Bakkali-Hassani_2023,Romero_theory} consisting of a majority and a minority component, leading to effective single-component self-focusing dynamics. In contrast, here we showcase the manifestation of features such as MI and rogue waves {\it for arbitrary particle ratios between the immiscible components}, all the way to equal atom mixtures. 
This constitutes a fundamental feature of this work and a key
digression from earlier studies, since MI dynamics indisputably
emerges even though our purely repulsively interacting balanced mixtures allow no reduction to an effectively attractive
system.
Excellent agreement among experiment and three-dimensional (3D) mean-field theory is observed. 
This setup enables a quantitative characterization of the emergent nonstationary, coherent, nonlinear oscillatory structures referred to as dispersive shock waves (DSWs), previously experimentally realized in repulsive single-component BECs~\cite{DSWPRAhoefer,DSWPRAhoefer2}, and it also enables the potential realization of collisions between them. 
The latter are experimentally achieved by utilizing a geometry of two repulsive barriers, mimicking a matter-wave cavity which triggers the interference of multiple dam breaks and leads to the formation of a PS~\cite{our2024PRL}. 

\begin{figure}[h!]
    \centering
    \includegraphics{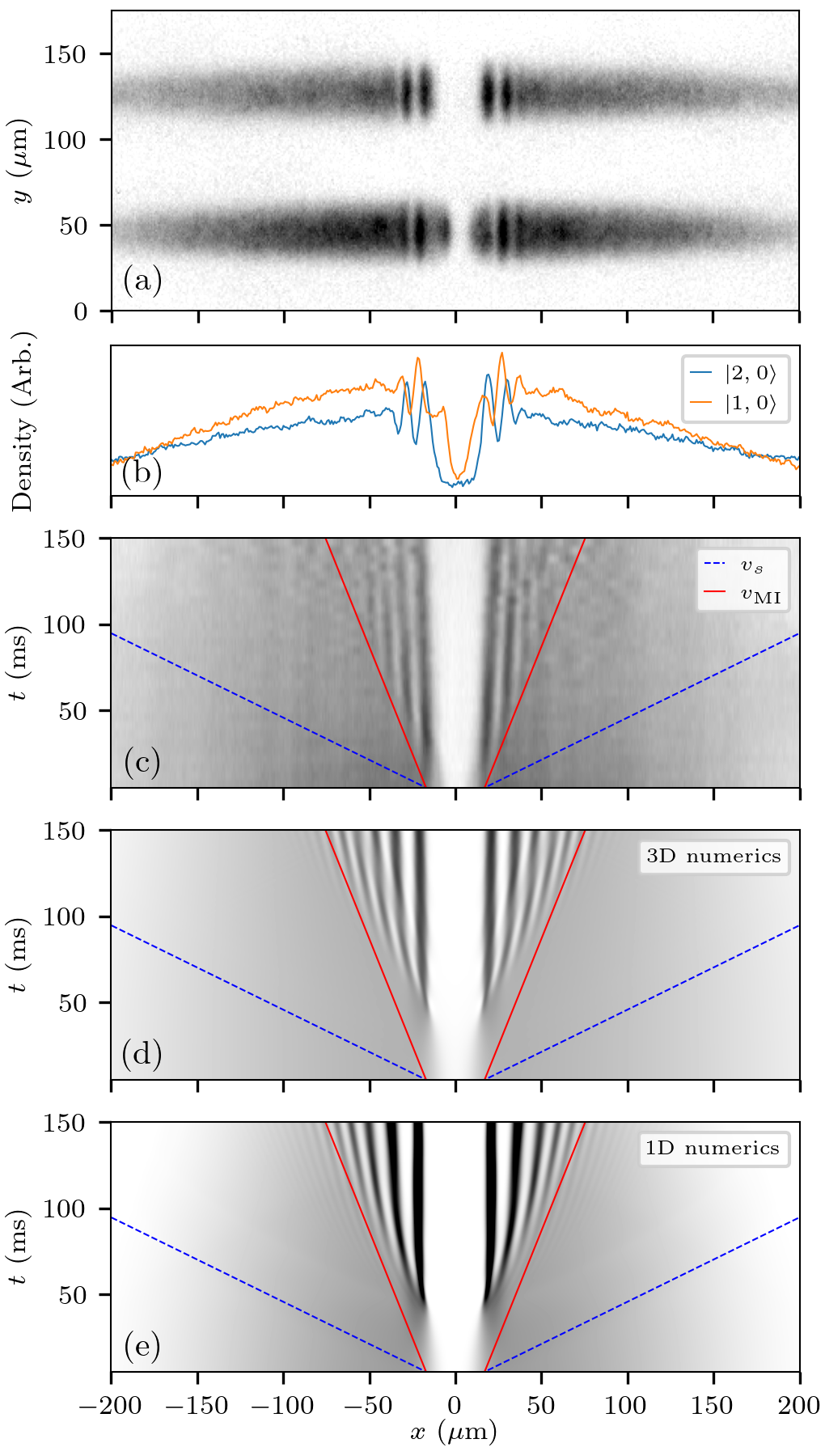}
    \caption{Experimental and numerical demonstration of the nonlinear stage of MI from a repulsive barrier for an initially 50:50 mixture of $\lvert2,0\rangle$ and $\lvert1,0\rangle$ atoms. 
    (a)~Single-shot absorption image taken after 60 ms of evolution time showing the $\lvert 2,0\rangle$ state above the $\lvert 1,0\rangle$ state.
    (b)~Integrated cross sections averaging over 20 independent experimental realizations corresponding to the absorption image in panel (a). 
    (c)~Spacetime evolution of the average integrated cross sections for the $\lvert2,0\rangle$ state. 
    The blue dashed lines show the experimentally determined speed of sound and the red lines depict the theoretically predicted MI expansion rate given in Eq.~\ref{eq:edge_gino}.
    Density evolution  of the (d) integrated cross sections from 3D numerical simulations and (e) 1D simulations of the experimental conditions overlaid with the same lines as in panel (c).}
    \label{fig:barrier_exp}
\end{figure}

\paragraph*{Experimental results.}
We begin with a BEC of approximately $9\times10^5$ $^{87}$Rb atoms in an elongated optical trap with frequencies $\omega_{x,y,z} = 2\pi \times (2.5, 246, 261)$~Hz.
The BEC is prepared in the $\lvert F, m_F\rangle = \lvert 1,-1 \rangle$ state in the presence of a narrow Gaussian beam generated by a 660~nm laser.
The beam produces a repulsive barrier for the atoms that is placed at the center of the trap, with widths $s_x = 7.5$ and $s_y = 59.8$~$\mu$m and a potential height corresponding to approximately $I_0 = 300$~nK, sufficient to completely split the condensate.
This initial state shows no meaningful dynamics over long timescales if left as a single spin component \cite{supplement}.
To initiate dynamics, we use fast microwave and rf pulses to prepare a mixture of the $\lvert 1,0 \rangle$ state with the $\lvert 2,0 \rangle$ state. 
This procedure results in a {\it highly reproducible initial state}, followed by MI features deterministically seeded by the edge of the stationary barrier. 
There is a slight atom loss with an exponential decay time of approximately 220 ms for the atoms in the $\lvert 2,0\rangle$ state, which, however, we have fully taken into account in the numerical simulations and find that it does not significantly impact the dynamics.

A principal experimental result of this work is summarized in Figs.~\ref{fig:barrier_exp}(a)-(c).
Fig.~\ref{fig:barrier_exp}(a) shows an example of a single experimental absorption image taken 60 ms after the preparation of an equally populated spin mixture, with the $\lvert 2,0\rangle$ state appearing as the upper cloud and $\lvert 1,0\rangle$ as the lower cloud in the panel. 
Each image is taken after 14 ms of free fall, allowing the $\lvert 2,0\rangle$ state to be transferred back to the $\lvert 1,-1\rangle$ state, followed by a Stern-Gerlach process to image the two spin components simultaneously. 
Fig.~\ref{fig:barrier_exp}(b) depicts the integrated cross sections corresponding to Fig.~\ref{fig:barrier_exp}(a) but averaged over 20 independent experimental realizations.
The complementarity of the density modulations in the two spin states demonstrates the energy scale of these excitations -- the instability excites spin oscillation modes without significantly affecting the overall condensate density. 
Moving forward, we present only the $\lvert 2,0\rangle$ state recognizing that the $\lvert 1,0\rangle$ state forms a complementary pattern.

The experimental procedure is repeated for 20 independent observations, taken every 5 ms from the creation of the spin mixture up to 150 ms later.
Each integrated cross section is then represented as a row on the space-time diagram in Fig.~\ref{fig:barrier_exp}(c). 
The structures observed here are hallmarks of the nonlinear stage of MI, consisting of a gradual spatial transition between a harmonic outer edge and a central solitonic edge on either side of the barrier in the form of a DSW structure.
In Fig.~\ref{fig:barrier_exp}(c), dashed blue lines represent the speed of sound in this system, while red lines mark the speed of the MI envelope predicted by the 1D analytical expression Eq.~\ref{eq:edge_gino} below. 
Good agreement between the expected MI growth rate and the experimental observations is found.
The speed of sound in this system was determined experimentally to be $v_s = 2.0(1)\ \mu$m/ms in a separate experiment, under similar conditions, where the barrier was suddenly doubled in strength to produce two outward traveling sound pulses, see also the supplemental material (SM)~\cite{supplement}. 
In addition, we simulate the experimental conditions described above in both a 3D and an effective 1D model (detailed in SM~\cite{supplement}, including Refs.~\cite{Salasnich2002, NICOLIN20104663}) shown in Figs.~\ref{fig:barrier_exp}(d) and \ref{fig:barrier_exp}(e), respectively.
In all cases, we observe excellent agreement with the experimental observations which are also in line with the analytical predictions of the MI growth. Intriguingly, these results also bear a striking resemblance to the experimental results of a single component with attractive interactions in a nonlinear optical fiber described by Kraych, et al.~\cite{Kraych2019}, even though we are working with a two-component system with purely repulsive interactions.

\paragraph*{Modulationally unstable dynamics.}

For the numerical implementation of our setup, we begin with all atoms residing in the $\lvert 1,-1 \rangle$ hyperfine state of $^{87}$Rb. 
As a first step, the mean-field ground state of this single-component setup is obtained, via imaginary time-propagation under the influence of the external potential of Eq.~(\ref{potential}) containing a repulsive barrier.
Next, we numerically simulate the experimental RF protocol by suddenly generating specific population mixtures, namely either a 50:50 or an 85:15 mixture of the $\lvert 1,0 \rangle$ and $\lvert 2,0 \rangle$ hyperfine states, respectively. 

The ensuing dynamics of the two-component system is monitored through the 
coupled 3D Gross-Pitaevskii equations (GPEs)~\cite{Pitaevskii2003,Kevrekidis2015} 
\begin{eqnarray}
i\hbar\frac{\partial\Psi_F}{\partial t} = \bigg[-\frac{\hbar^2}{2M} \nabla_{\textbf{r}}^2 + V(\textbf{r})
      + \sum_{F'=1}^2 g_{FF'}^{(3D)}|\Psi_{F'}|^2 \bigg] \Psi_F,
\label{eq:CGPE}
\end{eqnarray}
where $\textbf{r}=(x,y,z)$.
The wave function of the $F=1,2$ hyperfine state is $\Psi_F \equiv \Psi_F(\textbf{r},t)$. 
Additionally, $g_{FF'}^{(3D)}=4 \pi N_{F'} \hbar^2 a_{FF'}/M$ are the intra- $(F=F')$ and inter-component ($F\neq F'$) interaction coefficients related to the 3D s-wave scattering lengths $a_{FF'}$ (see below), $M$ is the mass of $^{87}$Rb atoms, whilst $N_F$ refers to the $F$ spin channel atom number.
The external potential consists of the harmonic trap and the superimposed optically induced repulsive potential barrier
\vspace*{-0.4ex}
\begin{eqnarray}
V(\textbf{r})= \frac12 M (\omega^2_x x^2 + \omega^2_y y^2 + \omega^2_z z^2 ) 
 + V_0\,e^{-2\bigg[\left(\frac{x}{s_x}\right)^2+\left(\frac{y}{s_y}\right)^2\bigg]}.\label{potential}
\end{eqnarray}
The potential barrier is characterized by its height $V_0$ and widths $(s_x,s_y)$, in line with the experimental values previously reported. 
This divides the BEC into two distinct, non-overlapping density regions.

The relevant scattering lengths are $a_{11}=100.86a_0$, $a_{22}=94.57a_0$, and $a_{12}=a_{21}=98.9a_0$, with $a_0$ denoting the Bohr radius. 
These scattering lengths result in an immiscible mixture.
In the case of a large population imbalance among the hyperfine states, they lead to 
effectively attractive
dynamics in the minority component, dictated by $a_\textnormal{eff}=a_{22}-a_{12}^2/a_{11}<0$~\cite{Dutton2005,Bakkali-Hassani2021,our2024PRL}. 
This, in turn, allows to observe focusing phenomena, such as the MI, in an otherwise repulsive environment. 
We emphasize that the use of the potential barrier is crucial for inducing the creation of experimentally repeatable and controllable counterpropagating DSWs.
Indeed, without the barrier, an immiscible mixture would be subject to spontaneous MI~\cite{supplement}.
Note also that the present protocol is in sharp contrast to the use of an attractive potential in Ref.~\cite{our2024PRL}, which facilitated rogue wave generation in a majority-minority scenario.

A typical example demonstrating the dynamical manifestation of MI and the resulting DSW formation is depicted in Fig.~\ref{fig:barrier_exp}(d). 
Here, our 3D mean-field simulations are based on the experimental parameters of  Fig.~\ref{fig:barrier_exp}(c) and account for an atom-loss rate in the $\lvert 2,0\rangle$ state.
The corresponding integrated density evolution, $n(x,t)=\int\,dy\,dz\,|\Psi_F(x,y,z;t)|^2$, reveals the emission of two counter propagating DSWs featuring density undulations, both stemming from the Riemann-type initial condition emulated by the steep barrier. 
These DSWs, which describe the nonlinear stage of MI, expand with a speed represented by the red lines in Fig.~\ref{fig:barrier_exp}(d). 
The relevant density modulations along with the characteristic speed of their envelope are more clearly discernible as compared to the experiment, see Figs.~\ref{fig:barrier_exp}(c), (d). 

\begin{figure}[t]
\centering
\includegraphics[width=\columnwidth]{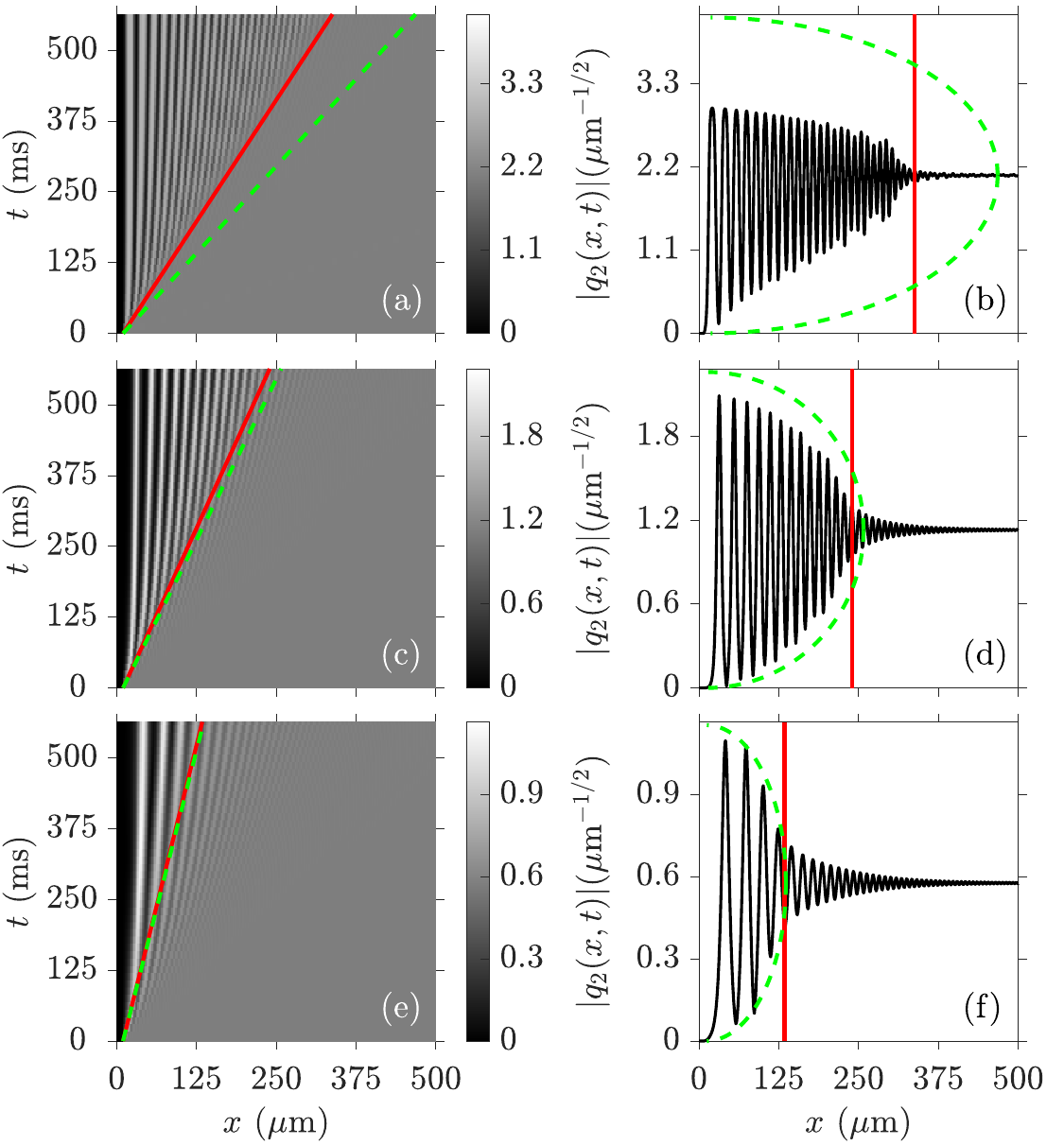}
\caption{Numerical 1D results of the nonlinear stage of MI of the $\lvert 2,0\rangle$ component in a (a)-(b) 50:50, (c)-(d) 85:15, and (e)-(f) 96:4 mixture. (a), (c), (e), Spatiotemporal evolution of the minority component's wavefunction magnitude, $|q_2(x,t)|$. 
(b), (d), (f), Snapshot of $|q_2(x,t)|$ at $t=564$~ms.
Red lines mark the edge of the modulated region described by Eq.~\eqref{eq:edge_gino}. 
Dashed green lines depict the envelope of the modulations dictated by Eq.~\eqref{eq:edge_gino_limit}. 
Due to symmetry, only the positive $x$-axis is shown.}
\label{fig:NLMI}
\end{figure}

\paragraph*{Nonlinear stage of MI.}\label{NL_stage}
To obtain a theoretical handle on the observed phenomenology, we start with the following coupled 1D NLS system, written in dimensionless units,
\begin{equation}
\label{eq:VNLS}
i q_{j,t} + q_{j,xx} - 2(g_{j1}|q_1|^2 + g_{j2}|q_2|^2)\,q_j = 0 \,.
\end{equation}
Here, $j=1,2$, while  constants $g_{jk}$ (for $j,k = 1,2$), quantify the effective 1D intra-  ($j=k$) and inter-component ($j\neq k$) interactions, and subscripts $t$ and $x$ denote temporal and spatial differentiation, respectively, of each component's 1D wavefunction $q_j(x,t)$.
It is known that for an immiscible mixture (i.e., $g_{11}g_{22} < g_{12}^2$), MI can occur even in repulsive, $g_{jk}>0$, BECs~\cite{Tommasini,Kasamatsu_MI1,Kasamatsu_MI2}.
On the other hand, to the best of our knowledge, the MI-driven dynamical response has never been analytically characterized nor experimentally observed in such systems.
It turns out that it is possible to prove (see SM~\cite{supplement} for the explicit derivation, including Ref.~\cite{NIST}) that in such a scenario, the speed of the edge of the modulated region generalizes to $x(t) = \pm V_\textnormal{2c}\,t$.
The involved velocity reads  
\begin{align}
\label{eq:edge_gino}
    V_\textnormal{2c} = 4\sqrt{- g_+ + \sqrt{g_-^2 + 4Q_1^2Q_2^2g_{12}^2}}\,,
\end{align}
with $g_\pm = g_{11} Q_1^2 \pm g_{22} Q_2^2$, and $Q_j$ denoting the background amplitude of each component.
The prediction of Eq.~\eqref{eq:edge_gino} can be compared to the analogous one of the single-component setting. 
The dynamics of the latter, characterized by the complex wavefunction $q(x,t)$, is governed by the NLS equation, $i q_t + q_{xx} - 2g|q|^2q = 0$, with $g<0$ ($g>0$) referring to the focusing (defocusing) nonlinearity. 
In this context, the edge of the modulated region propagates as $x(t) = \pm V_\textnormal{nls} t$~\cite{EL1993357,GinoDion}, having velocity$V_\textnormal{nls} = 4q_0\sqrt{-2g}$ (for background amplitude $q_0$).

For a highly particle-imbalanced two-component mixture, it is natural to take the limiting case of $Q_2\ll Q_1$ in Eq.~\eqref{eq:edge_gino} in order to identify analogies with the single-component scenario (of the minority component). 
Expanding the inner square root and neglecting higher-order terms leads to 
\begin{equation}
\label{eq:edge_gino_limit}
    V_\textnormal{2c} = 4 Q_2\sqrt{-2g_{\rm eff}} + \mathcal{O}(Q_2^2)\,.
\end{equation}
Eq.~\eqref{eq:edge_gino_limit} is equivalent to $V_\textnormal{nls}$ with an effective 1D coupling $g_\textnormal{eff} = g_{22} - g_{12}^2/g_{11}$ for the minority component of background amplitude $Q_2$.
This is in line with the earlier analysis of the effective single-component (minority) reduction in~\cite{Dutton2005,Bakkali-Hassani2021}.
However, we emphasize that Eq.~\eqref{eq:edge_gino} is valid in the more general case of arbitrary particle imbalanced BEC components. 

\begin{figure}[t]
\centering
\includegraphics{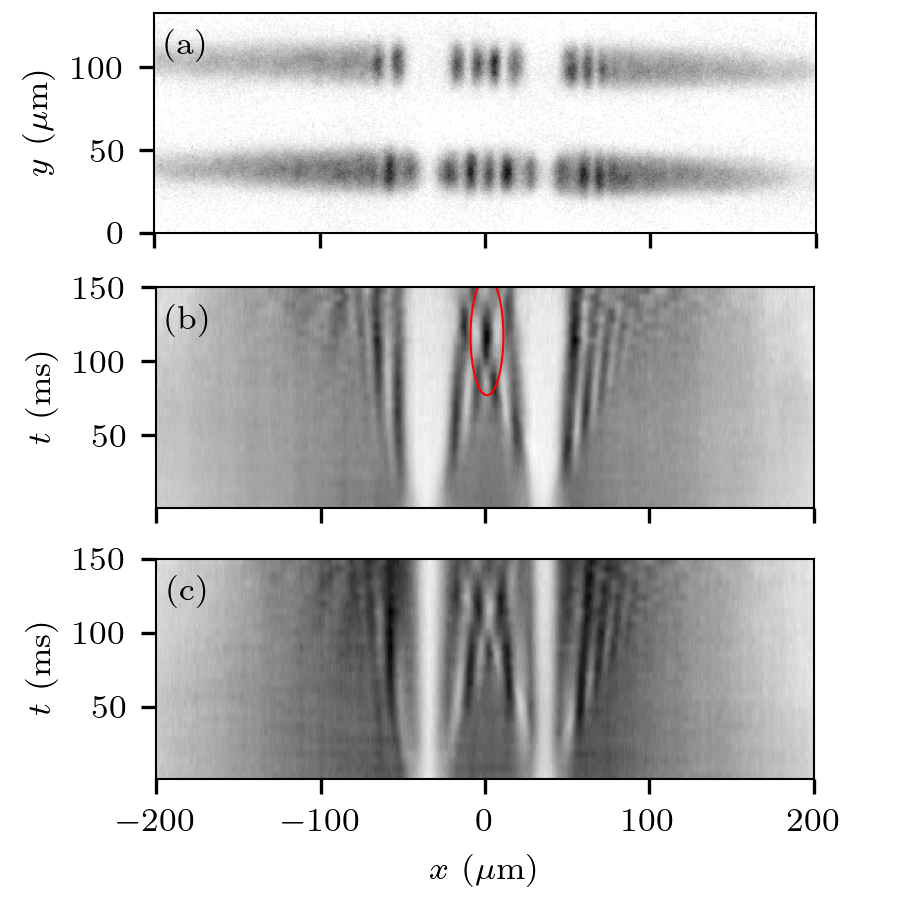}
\caption{Experimental images depicting DSW interaction produced by two barriers and subsequent PS  formation in a 50:50 immiscible mixture. 
(a) Absorption image taken after 80 ms of evolution time with $\lvert 2,0\rangle$ appearing above $\lvert 1,0\rangle$. 
Spacetime evolution of the (b) $\lvert 2,0 \rangle$ and (c) $\lvert 1,0 \rangle$   component featuring complementary density profiles. A PS (red circle) appears at the center of the interference region.}
\label{fig:dual_barrier}
\end{figure}

The resultant density dynamics of one of the components of the 50:50 mixture stemming from the 1D GPE analysis is depicted in Fig.~\ref{fig:barrier_exp}(e). 
Here, both the velocity of the MI edge, predicted by Eq.~(\ref{eq:edge_gino}), and that of the experimentally determined speed of sound are depicted. 
Evidently, the observed density modulations exhibit a propagation speed that is in accordance with the analytical prediction for the 50:50 mixture. 
Additionally, the transition between particle balanced to highly imbalanced settings (and hence effectively single-component) is demonstrated in Fig.~\ref{fig:NLMI}. 
Particularly, Figs.~\ref{fig:NLMI}(a), \ref{fig:NLMI}(c) and \ref{fig:NLMI}(e) illustrate the spatiotemporal evolution of the minority component's wavefunction magnitude, $|q_2(x,t)|$, accompanied by the edges of the MI in two- (solid red line) and single-component (green dashed line) setups [see Eqs.~\eqref{eq:edge_gino} and  \eqref{eq:edge_gino_limit}, respectively].
Additionally, Figs.~\ref{fig:NLMI}(b), \ref{fig:NLMI}(d) and \ref{fig:NLMI}(f) showcase specific snapshots at $t=564$~ms. 
Here, for reference, the envelope of the MI calculated from the single-component setting reported in Ref.~\cite{PRE2018v98p052220} is also depicted (see more details in SM~\cite{supplement}).
It can be readily seen that, as the particle imbalance increases, the MI occurs at longer evolution times and the envelope of the resulting structure expands more slowly. 
Interestingly, the analytical estimates of the MI speed are in agreement with the outcome of the mean-field simulations. 
The MI speed and envelope profile approach their single-component counterpart with $g=g_\textnormal{eff}$ in the limit of $Q_2\ll Q_1$, corroborating the relevant reduction. 

\paragraph*{Two-component dam-break problem.} 

Finally, we extend our considerations to a geometry consisting of two barriers, each of which produces density modulations on either of its sides, as presented in Fig.~\ref{fig:dual_barrier}. 
This more exotic scenario, resembling a matter-wave cavity (e.g., motivated by similar formulations in the case of solitonic collisions~\cite{Lannig}), produces an interference pattern of the two DSWs emerging between the barriers as a result of the nonlinear stage of MI. 
An important byproduct of such interference in the semi-classical limit of focusing NLS models has been theoretically identified relatively recently in the work of Ref.~\cite{NLTY2016v29p2798}.
There, it was found that this interference results in the formation of PS-like structures, a feature that was justified based on the analysis of the integrable properties of the relevant model. 
A realization of this recently proposed mechanism was reported
in optics~\cite{Audo:18}, but never in multiple components or in any
form in quantum fluids, to the best of our knowledge.
Here, we showcase the fact that even in our two-component, repulsive interaction generalization, both the numerical simulations \cite{supplement}, but crucially also the experimental data of Figs.~\ref{fig:dual_barrier}(b) and  \ref{fig:dual_barrier}(c) support the (unprecedented, to our knowledge) manifestation of this type of PS-generating dam-break interference in the realm of atomic BECs. 
Remarkably, this is true even for an equal mixture.
In fact, we have further verified the existence of a PS-like structure by 
comparing its properties with those of the analytically available Peregrine waveform~\cite{peregrine1983water,Romero_theory}. 
This reveals a very good agreement within the region of the PS-like core accompanied by a $\pi$ phase jump between its core and tails, see also~\cite{supplement} for details. 
Also, we have checked experimentally that an increasing inter-barrier distance leads to PS-like formation at longer times. 

\paragraph*{Conclusions.}

We have examined the nonlinear stage of MI in a two-component repulsive BEC of two hyperfine states of $^{87}$Rb atoms, featuring a repulsive barrier.
The manifestation of MI, which leads to DSW formation, is experimentally observed for immiscible species, and theoretically analyzed for arbitrary ratios between the two components. 
Expanding on the analysis of the reducible majority-minority case, we favorably compare our experimental and numerical results with analytical formulae for the relevant wavefunction, its envelope, and propagation speed.
Beyond the very good agreement between 3D numerics and experimental observations, and the definitive qualitative understanding enabled by the 1D analysis, we have also explored the more exotic scenario where two repulsive barriers, emulating a matter-wave cavity, induce the interference of multiple dam-breaks.
These, in line with relevant earlier theoretical predictions~\cite{NLTY2016v29p2798}, have been verified to seed the emergence of a PS structure, providing an intriguing alternative avenue for its experimental realization compared to the gradient catastrophe one~\cite{bertola} recently realized within BECs~\cite{our2024PRL}.

This work provides exciting avenues for further exploration, including a quantitative characterization of the nonlinear stage of MI for arbitrary coefficients in multi-component systems utilizing Whitham modulation theory~\cite{PRE2018v98p052220,EL201611}, or the investigation of the impact of spinor settings or progressively higher dimensions, which may give rise to similar or other interesting patterns~\cite{PhysRevLett.125.250401}. 
The identification of underlying correlation phenomena~\cite{Mistakidis_review} emanating from the nonlinear stage of MI is of considerable interest. 
Another intriguing prospect is the study of possible  universal features when the barrier is driven, triggering  turbulent cascades~\cite{navon2016emergence,navon2021quantum}. 

\paragraph*{Acknowledgements.} 

A.R.R. acknowledges MCIN/AEI/10.13039/501100011033 from grants: PID2023-147112NB-C22; CNS2022-135529 through the “European Union NextGenerationEU/PRTR”; CEX2019-000918-M through the “Unit of Excellence Mar\'ia de Maeztu 2020-2023” award to the Institute of Cosmos Sciences; and by the Generalitat de Catalunya, grant 2021SGR01095.
P.E. acknowledges funding from NSF through Grant No. PHY-2207588 as well as support from a Boeing Endowed Professorship at WSU. This research was also supported in part by the National Science Foundation under Grant No. NSF PHY-1748958.
S.I.M. acknowledges support from the Missouri Science and Technology, Department of Physics, Startup fund.  
This material is also based upon work supported by the US
National Science Foundation under Grant No. PHY-2110030,
PHY-2408988 and DMS-2204702 (P.G.K.).

\onecolumngrid
\clearpage

\renewcommand{\thefigure}{S\arabic{figure}}
\renewcommand{\theequation}{S\arabic{equation}}
\setcounter{equation}{0}
\setcounter{figure}{0}

\begin{center}
	{\Large\bfseries Supplementary Material: Nonlinear Stage of Modulational Instability\\ in Repulsive Two-Component Bose-Einstein Condensates \\ 
 }
\end{center}

\bigskip
\section{Experimental Setup}

Our experiments begin with a combination of laser cooling and evaporative cooling resulting in a Bose-Einstein condensate (BEC) composed of $^{87}$Rb atoms prepared in the  $\lvert F, m_F\rangle = \lvert 1,-1\rangle$ hyperfine state. 
The atoms are supported against gravity in an optical dipole trap formed by a single, tightly focused laser beam with a wavelength of 1064 nm that propagates horizontally. 
This optical trap is state independent, simultaneously trapping different hyperfine states in an elongated harmonic potential with trap frequencies of $(\omega_x,\omega_y,\omega_z) = 2\pi \times (2.5, 246, 261)$ Hz. 

To generate a repulsive barrier at the center of the cloud, a second beam with wavelength 660 nm is turned on in the vertical direction, splitting the BEC into two halves along its elongated axis.
Similarly to the attractive dipole trap beam, this wavelength is sufficiently detuned from atomic resonances to produce a state-independent potential, and in particular blue detuned resulting in a repulsive barrier.
Next, a homogeneous magnetic bias field of 10~G is applied that separates the energy levels of the hyperfine states, allowing us to later prepare arbitrary mixtures of states by driving radio frequency (RF) transitions (within the $F=1$ manifold or the $F=2$ manifold) or microwave (MW) transitions (between the $F=1$ and $F=2$ manifolds). 
The BEC is allowed to equilibrate in the presence of the magnetic field and the repulsive barrier for 2 seconds, before proceeding with the state preparations. At this point, the BEC contains approximately $9\times10^5$ atoms.

To prepare a 50:50 population mixture of atoms in the $\lvert 1,0\rangle$ and $\lvert 2,0\rangle$ states, we first apply a 51~$\mu$s RF pulse on the $\lvert 1,-1\rangle$ to $\lvert 1,0\rangle$ transition to transfer 50\% of the atoms from the $\lvert 1,-1\rangle$ state to the $\lvert 1,0\rangle$ state. 
This is followed by a 2~ms MW frequency sweep across the $\lvert 1,-1\rangle$ to $\lvert 2,0\rangle$ transition to transfer all remaining $\lvert 1,-1\rangle$ atoms to the $\lvert 2,0\rangle$ state using an adiabatic rapid passage.   
The preparation of the state mixture forms the starting point for the evolution described in the main text. 
During this evolution, both components remain trapped in the optical confinement described above.

\begin{figure}[b]
\includegraphics{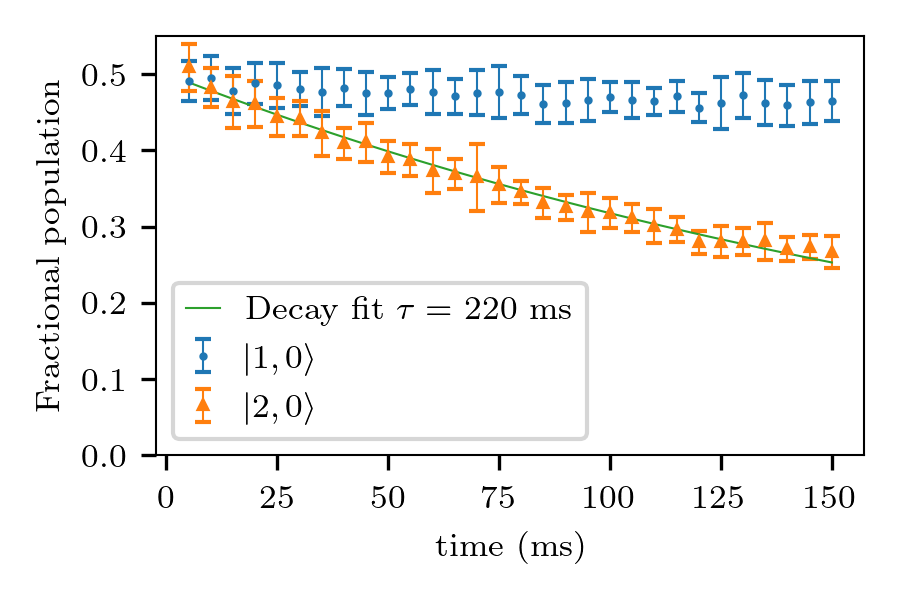}
\caption{Fractional populations of the two spin populations over time corresponding to the data shown in Fig.~1 of the main text. An exponential fit to the decay in the $\ket{2,0}$ population is shown to have a time constant of $\tau=220$~ms.}
\label{fig:sup_atom_number}
\end{figure}

During the evolution, some loss of population occurs in the $F=2$ spin manifold due to spin-changing collisions.
Fig.~\ref{fig:sup_atom_number} shows the experimentally observed population fraction of each spin state for the data used in Fig.~1 of the main text, normalized to the initial 50:50 populations.
The atom loss rate is phenomenologically captured by an exponential fit with decay time of approximately $\tau = 220$ ms and then incorporated into the numerical simulations as a time dependent loss rate in the $F=2$ population.

To image the atomic cloud, we first transfer all atoms that are in the $\lvert 2,0\rangle$ state back into the $\lvert 1,-1\rangle$ state using again a 2~ms MW  sweep.
The optical trap is removed and a Stern-Gerlach procedure is employed to separate the two hyperfine states vertically as they fall, allowing the two spin populations to be imaged at separate regions of the CCD camera. 
This procedure relies on the fact that atoms in the $\lvert 1,-1\rangle$ or $\lvert 1,0\rangle$ state have different magnetic moments. 
Therefore, a brief application of a vertical magnetic gradient leads to a differential force on the two states so that they separate vertically in a short time-of-flight period. 
Images are then acquired using a 10~$\mu$s long illumination pulse on the $F=1$ to $F'=1$ transition. This imaging procedure is destructive, requiring repeated runs of the experiment to produce a time sequence.
The spontaneous MI observed at longer times is irreproducible, however, the barrier induced dynamics is highly reproducible and thus the time evolution can fully be studied even using a succession of experimental runs with destructive imaging.

\section{One-dimensional reduction}
We begin by illustrating the correspondence between the three-dimensional (3D) Gross-Pitaevskii equation (GPE),
\begin{equation}
    i\hbar\frac{\partial\Psi}{\partial t} = \left[-\frac{\hbar^2}{2M} \nabla_{\textbf{r}}^2 + V(\textbf{r})
    + g^{\rm (3D)}|\Psi|^2 \right] \Psi \,,
\label{eq:GPE}    
\end{equation}
and the one-dimensional (1D) nonlinear Schr\"odinger (NLS) equation,
\begin{equation}
    i q_t + q_{xx} - 2g|q|^2q = 0 \,.
\label{eq:NLS}    
\end{equation}
For simplicity, we focus on the single component case, but the following discussion can be easily extended to multicomponent systems.

In order to achieve  effective 1D dynamics, it is important to avoid excitations in the transverse directions. 
This can be accomplished by employing a tight transverse  cigar-shaped trapping potential, i.e., $\omega_x\ll\omega_\perp$, with $a_\perp < \xi$, or equivalently $\hbar\omega_\perp > \mu$.
Here, $a_\perp = \sqrt{\hbar/M\omega_\perp}$ is the transverse harmonic oscillator length, $\xi$ denotes the healing length of the Bose gas, and $\mu$ is the chemical potential.
Thus, the 3D wavefuntion, $\Psi(\textbf{r},t)$, can be decomposed into longitudinal, $\psi(x,t)$, and transverse, $\chi(y,z)$, degrees-of-freedom, namely $\Psi(\textbf{r},t) = \psi(x,t)\chi(y,z)e^{-i\mu t/\hbar}$.
Subsequently, integrating out the transverse degrees-of-freedom~\cite{Pitaevskii2003}, we obtain the 1D GPE,
\begin{equation}
    i\hbar\frac{\partial\psi}{\partial t} = \left[-\frac{\hbar^2}{2M} \frac{\partial^2}{\partial x^2}+ V(x) + g^\textrm{(1D)}|\psi|^2 \right] \psi \,,
\label{eq:GPE_1D}    
\end{equation}
where $V(x)=M\omega_xx^2/2$ and $g^{(1D)}$ denotes an effective 1D interaction strength.
Lastly, with the change of variables $x \to a_\perp x'$, $t \to \omega_\perp^{-1} t'$, and $\psi \to a_\perp^{-1/2}\psi'$, Eq.~\eqref{eq:GPE_1D} becomes nondimensional,
\begin{equation}
    i\frac{\partial \psi'}{\partial t'} = \left[-\frac{1}{2} \frac{\partial^2}{\partial x'^2}+ V'(x') + g'|\psi'|^2 \right] \psi' \,,
\label{eq:GPE_1D_non}    
\end{equation}
with $V'(x') = V(x')a_\perp^2/\hbar\omega_\perp$, and $g'=g^{(1D)}/\hbar\omega_\perp a_\perp$.
Apparently, if $V'(x') = 0$ and $t' \to t'/2$, we recover the NLS Eq.~\eqref{eq:NLS}.

Here, we would also like to emphasize that, typically, an expression for $g^{\rm (1D)}$ is obtained when integrating out the transverse degrees-of-freedom.
In order to do so, one needs to solve an auxiliary problem under suitable assumptions.
For instance, Eq.~\eqref{eq:GPE_1D} can be obtained by assuming that the solution of the transverse degrees-of-freedom corresponds (in a suitably dilute limit) to a Gaussian profile~\cite{Salasnich2002}.
If instead it is assumed that they correspond to a Thomas-Fermi profile, one obtains a different model derived in~\cite{NICOLIN20104663}.
Here, despite achieving effective 1D dynamics, the transverse degrees-of-freedom do not totally match either of the aforementioned assumptions, and thus we opted to identify the effective $g^{\rm (1D)}$ using a more empirical approach to match the experiment, keeping in mind the fact that $\int |\Psi(\textbf{r},t)|^2 \dd\textbf{r} = 1$, and subsequently $\int |\chi(y,z,t)|^2 \dd y\,\dd z = 1$ and $\int |\psi(x,t)|^2 \dd x = 1$.
More specifically, one can solve the time-independent 1D GPE for a chosen $g^{(\textnormal{1D})}$ under the constraint that both the numerical and experimental radius of the BEC must coincide.
However, if $g^{(\textnormal{1D})}$ has not been obtained through the above reduction, the last of the three aforementioned equalities will not be automatically fulfilled. 
Hence, we identify $\eta$ such that $\int |\psi(x,t)|^2 \dd x =  \eta$ instead.
Here, we interpret $\eta$ as a corrective coefficient to the assumptions taken to solve the auxiliary problem.
Then, by inferring on the form of the interaction term in Eq.~\eqref{eq:GPE_1D}, we can redefine $g^{(\textnormal{1D})}\to g_{\rm eff}^{(\textnormal{1D})}=\eta g^{(\textnormal{1D})}$, so that $\int |\psi(x,t)|^2 \dd x = 1$.

\section{Linearized stability analysis of the coupled NLS system}

Below, we present the linearized stability analysis for the background solution of a general coupled system of NLS equations for the field $\textbf{q}(x,t) = (q_1(x,t),q_2(x,t))^T$, which read 
\vspace*{-1ex}
\bse
\label{e:CNLS}
\begin{gather}
i q_{1,t} + q_{1,xx} - 2 (g_{11} |q_1|^2 + g_{12}|q_2|^2)\,q_1 = 0\,,
\\
i q_{2,t} + q_{2,xx} - 2 (g_{21} |q_1|^2 + g_{22}|q_2|^2)\,q_2 = 0\,.
\end{gather}
\ese
The relevant analysis has been provided earlier (see, as a relevant example,~\cite{Tommasini,Kasamatsu_MI1}), yet we include it here for reasons of completeness.

This system of equations admits the uniform background solution $\textbf{q}_o(x,t) = (Q_1\,e^{i\mu_1 t},Q_2\,e^{i\mu_2 t})$, with $Q_1$ and $Q_2$ being arbitrary real constants (which can be taken to be non-negative thanks to the phase invariance of the coupled NLS system~\eqref{e:CNLS} without loss of generality).
Accordingly, from the stationary equations of motion we obtain the chemical potentials of each component $\mu_i = 2(g_{ii}Q_i^2 + g_{ij}Q_j^2)$ with $i \neq j=1,2$.
It is convenient to perform a trivial phase rotation and consider the modified field $\textbf{q}(x,t) = (q_1(x,t), q_2(x,t))^T$ with $\tilde{q}_j(x,t) = q_j(x,t)\,e^{-i\mu_jt}$, $j=1,2$, which satisfies the NLS equations
\bse
\label{e:CNLS2}
\begin{gather}
i \tilde{q}_{1,t} + \tilde{q}_{1,xx} - 2 (g_{11} |\tilde{q}_1|^2 + g_{12}|\tilde{q}_2|^2 - \mu_1)\,\tilde{q}_1 = 0\,,
\\
i \tilde{q}_{2,t} + \tilde{q}_{2,xx} - 2 (g_{21} |\tilde{q}_1|^2 + g_{22}|\tilde{q}_2|^2 - \mu_2 )\,\tilde{q}_2 = 0\,.
\end{gather}
\ese
It is evident that these admit the constant solution $\textbf{q}_o(x,t) = (Q_1,Q_2)$.
The reason for considering a system with uniform background and no potential is twofold: 
(i) this is the simplest setting that gives rise to the phenomena in question (MI-induced DSW); 
(ii) the results of the analysis with no potential and a uniform background are also applicable to the case of a sharp spatial transition between zero and non-zero density background (e.g., see Ref.~\cite{PRE2018v98p052220}).

Next, we search for solutions representing a small perturbation of the constant background, namely $\tilde q_j(x,t) = Q_j(1+ w_j(x,t))$  with $|w_j(x,t)|\ll1$, $j=1,2$.
Substituting this ansatz into Eqs.~\eqref{e:CNLS2} and neglecting quadratic and higher-order terms we arrive to the linearized coupled NLS system
\bse
\begin{gather}
i w_{1,t}+w_{1,xx} - 2 Q_1^2 g_{11}(w_1 + w_1^*) - 2 Q_2^2 g_{12} (w_2 + w_2^*) = 0,
\\
i w_{2,t}+w_{2,xx} - 2 Q_1^2 g_{21}(w_1 + w_1^*) - 2 Q_2^2 g_{22}(w_2 + w_2^*) = 0\,,
\end{gather}
\ese
where the asterisk denotes complex conjugation.
Subsequently, it is possible to rewrite the above system into four individual evolution equations for the real and the imaginary components, by letting $w_j(x,t) = u_j(x,t) + iv_j(x,t)$ with $u_j(x,t)$ and $v_j(x,t)$ being  real-valued, for $j=1,2$. 
As such, we obtain the following system
\bse
\begin{gather}
 u_{1,t} + v_{1,xx} = 0 , \\
 u_{2,t} + v_{2,xx} = 0 , \\
 v_{1,t} - u_{1,xx} + 4 Q_1^2 g_{11} u_1 + 4 Q_2^2 g_{12} u_2  = 0 , \\
 v_{2,t} - u_{2,xx} + 4 Q_1^2 g_{21} u_1 + 4 Q_2^2 g_{22} u_2  = 0 .
\end{gather}
\ese
Assuming plane-wave solutions of the above system in the form $(u_j(x,t),v_j(x,t)) = (U_j,V_j)\,e^{i(kx - \omega t)}$, $j=1,2$,\label{disp_relation} we can derive the linear homogeneous system of equations $M \textbf{U} = 0$, 
where $\textbf{U} = (U_1,U_2,V_1,V_2)^T$ and the coefficient matrix $M$ reads 
\be
M = \begin{pmatrix} k^2 + 4 g_{11} Q_1^2  & 4 g_{21} Q_1^2  & i \omega  & 0 \\
 4 g_{12} Q_2^2 & k^2 + 4 g_{22} Q_2^2 & 0 & i \omega  \\
 -i \omega  & 0 & k^2 & 0 \\
 0 & -i \omega  & 0 & k^2 \\
\end{pmatrix}.
\ee
Apparently, nontrivial solutions are obtained when $\det (M) = 0$. 
This yields two different branches for the linearized dispersion relation 
\bse
\be
\omega_\pm ^2 = k^2 \bigg(k^2 + 2 (g_{11} Q_1^2 +  g_{22} Q_2^2) \pm 2 \sqrt{\Delta} \, \bigg),
\label{linearizeddispersionrelation}
\ee
where the interaction and amplitude dependent parameter 
\be
\Delta = g_{11}^2 Q_1^4 + g_{22}^2 Q_2^4 + 2 Q_1^2 Q_2^2 \left( 2 g_{12}g_{21} - g_{11}g_{22} \right) \,.\label{Delta_MI_parameter}
\ee
\ese
It is worthwhile to mention that this dispersion relation holds for arbitrary particle and mass imbalanced repulsive two-component Bose gases.

\section{Speed of sound and MI condition}

The solution branch obtained from $\omega_+^2$ given by Eq.~(\ref{linearizeddispersionrelation}) yields sound waves in the limit $k\to0$.
Indeed, it is straightforward to verify that, for this branch, the linearized group velocity associated with the branch \textcolor{blue}{$\omega_+(k)$} of the dispersion relation, $c_g(k) = d\omega_+/dk$, at $k=0$ takes the form
\be
c_g(0) = \sqrt2 \sqrt{g_{11} Q_1^2 +  g_{22} Q_2^2 + {\color{blue}\sqrt{\Delta}}}\,.\label{eq:analytic_sound_speed}
\ee
In the limiting (integrable) case where $g_{i,j} =1$ and the amplitudes $Q_1$ and $Q_2$ are normalized so that $Q_1^2+Q_2^2=1$, one simply obtains $c_g(0) = 2$.

\begin{figure}[h!]
    \centering
    \includegraphics{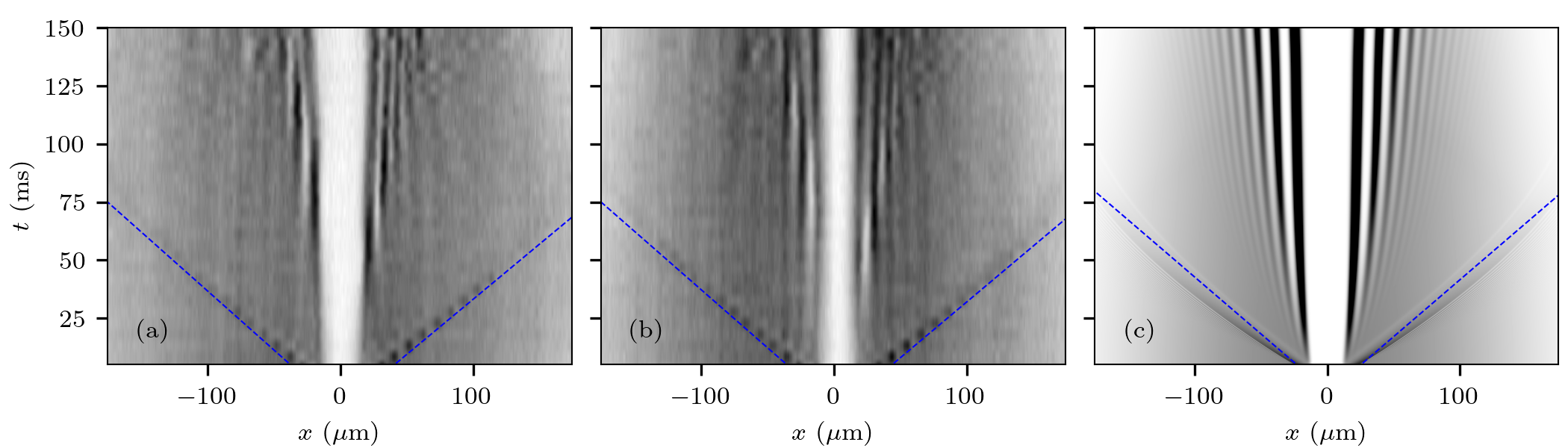}
    \caption{Experimental and numerical determination of the speed of sound. An equal spin mixture of (a) $\lvert 2,0\rangle$ and (b) $\lvert 1,0\rangle$ is prepared in the presence of an optical barrier. A sudden increase of the barrier height is used to induce sound waves. Each panel is an average over 10 experimental realizations for every 5 ms. Blue lines mark the linear fit to the traveling sound pulses from the experiment used to determine the speed of sound, averaging at $v_s = 2.0(1)\ \mu$m/ms. (c) 1D GPE simulations of the experimental procedure showing a consistent result for the speed of sound compared to the average over the experimental results, shown as blue lines.}
    \label{fig:soundwaves}
\end{figure}

This branch corresponds to the normal speed of sound in the fluid. 
As a consistency check, we determine this speed of sound in the context of the 1D GPE simulations as well as in the experiment.
In both cases, we prepare an equally populated spin mixture of $\lvert 2,0\rangle$ and $\lvert 1,0\rangle$ states in the presence of a strong optical barrier as described in the main text.
We then double the barrier's strength, inducing small amplitude sound waves traveling outward from the edges of the barrier.
Fig.~\ref{fig:soundwaves} shows the waves traveling outward on a spacetime diagram, from which we can then extract the speed of sound in this system. 
Our experimental measurements, corroborate the numerics and provide an estimate of the speed of sound of $v_s = 2.0(1)\ \mu$m/ms. 
This result is consistent with the speed of sound determined by the analytical result in Eq.~\ref{eq:analytic_sound_speed}, at $v_s = 2.089\ \mu$m/ms.

On the other hand, the branch obtained from $\omega_-^2$ explicates that MI is present when $\omega_-^2<0$ for some range of values of the wavenumber $k$. 
This concurrently entails an exponential growth of the perturbations.
Explicitly, this occurs when the constant term in the parenthesis of Eq.~\eqref{linearizeddispersionrelation} is negative, i.e., $2 (g_{11} Q_1^2 +  g_{22} Q_2^2) < \mp \sqrt{\Delta}$, which in turn, immediately implies that MI takes place only for $\omega_-$.
Simple re-arrangements of the above conditions then show that
a necessary and sufficient condition for the existence of MI is $g_{11}g_{22} - g_{12}g_{21}<0$, namely that the mixture is immiscible. 
Notice that this result is consistent with the analysis of Ref.~\cite{Tommasini,Kasamatsu_MI1}. 
In other words, even when all coupling terms are individually defocusing, MI still occurs if the cross-coupling terms dominate
compared to the self-coupling ones.

When MI is present, further inspection of Eq.~\eqref{linearizeddispersionrelation} determines the range of unstable wavenumbers, i.e., $|k|<k_\textnormal{max}$, with
\be
k_\textnormal{max}^2 =2\left(\sqrt\Delta - (g_{11} Q_1^2 +  g_{22} Q_2^2)\right)\,.
\ee
It is straightforward to check that in the respective scalar reduction, the above expression agrees with established results. 
For instance, in the case of $Q_2=0$ and $g_{11}=1$ (i.e., in the scalar defocusing scenario), one obtains $k_\textnormal{max}=0$ (implying absence of MI), whereas if $Q_2=0$ but $g_{11}=-1$ (i.e., in the scalar focusing case), one recovers $k_\textnormal{max} = 2 Q_1$ (e.g., see~\cite{EL1993357,GinoDion}).

\section{Expansion rate of the MI wedge}

Importantly, the linearized stability analysis presented above also allows one to obtain a prediction for the expansion rate of the MI wedge, as shown in Fig.~1 of the main text.  
Recall that, in a homogeneous one-component setting the propagation speed of the boundary of the DSW is the {\it minimum} of the linearized group velocity~\cite{EL1993357,GinoDion,SIREV2018v60p888}.  
The same is true in our case.
One can gain an intuitive understanding of this by considering how the various wave modes travel out from the narrow region of curvature near the central repulsive barrier.  
The sharp density gradient results in the excitation of a broad range of wavenumbers.
Excitations corresponding to stable wavenumbers propagate away with the linearized group speed, and therefore do so with a speed larger than the minimum of the group speed.
Therefore, the spatial extent of the DSW is confined to the region left over ``behind'', once the (stable) linear waves propagate away. 
Conversely, excitations corresponding to unstable wavenumbers grow without traveling producing the expanding DSW amplitude pattern.
\newpage

In light of the above discussion, the edges of the modulated region are given by the minimum of the linearized group velocity associated with the branch $\omega_-(k)$ of the dispersion relation.
In particular, from Eq.~\eqref{linearizeddispersionrelation} we find the linearized group velocity $c_g(k) = d\omega_-/dk$ as 
\be
c_g(k) = 2\frac{k^2 + g_{11} Q_1^2 + g_{22} Q_2^2 - \sqrt{\Delta}}
  {\sqrt{k^2 + 2 (g_{11} Q_1^2 + g_{22} Q_2^2 ) - 2 \sqrt{\Delta}}}\,,
\label{e:cg}
\ee
with $\Delta$ as in Eq.~\eqref{Delta_MI_parameter}.
The minimum of the above expression is found to occur when $k= k_\textnormal{min}$ with
\be
k_\textnormal{min} = \sqrt{3} \sqrt{- g_{11} Q_1^2 - g_{22} Q_2^2 + \sqrt{\Delta}}.
\ee
The corresponding minimum value of the linearized group velocity~\eqref{e:cg} reads 
\be
c_{g,\textnormal{min}} = 4 \sqrt{ - g_{11} Q_1^2 - g_{22} Q_2^2 + \sqrt{\Delta}}\,. \label{minimum_lin_vel}
\ee
This yields Eq.~4 in the main text.
As also discussed in the main text, Eq.~(\ref{minimum_lin_vel}) reduces to the well-known prediction for the NLS equation in the scalar reduction.
On the other hand, Eq.~(\ref{minimum_lin_vel}) remains valid for arbitrary values of the ratio of the amplitudes of the two components.
This result is a key outcome of our theory which can be readily compared against the direct numerical simulations, as well as the experimental observations, as is done in the main text.

\begin{figure}[b!]
    \centering
    \includegraphics{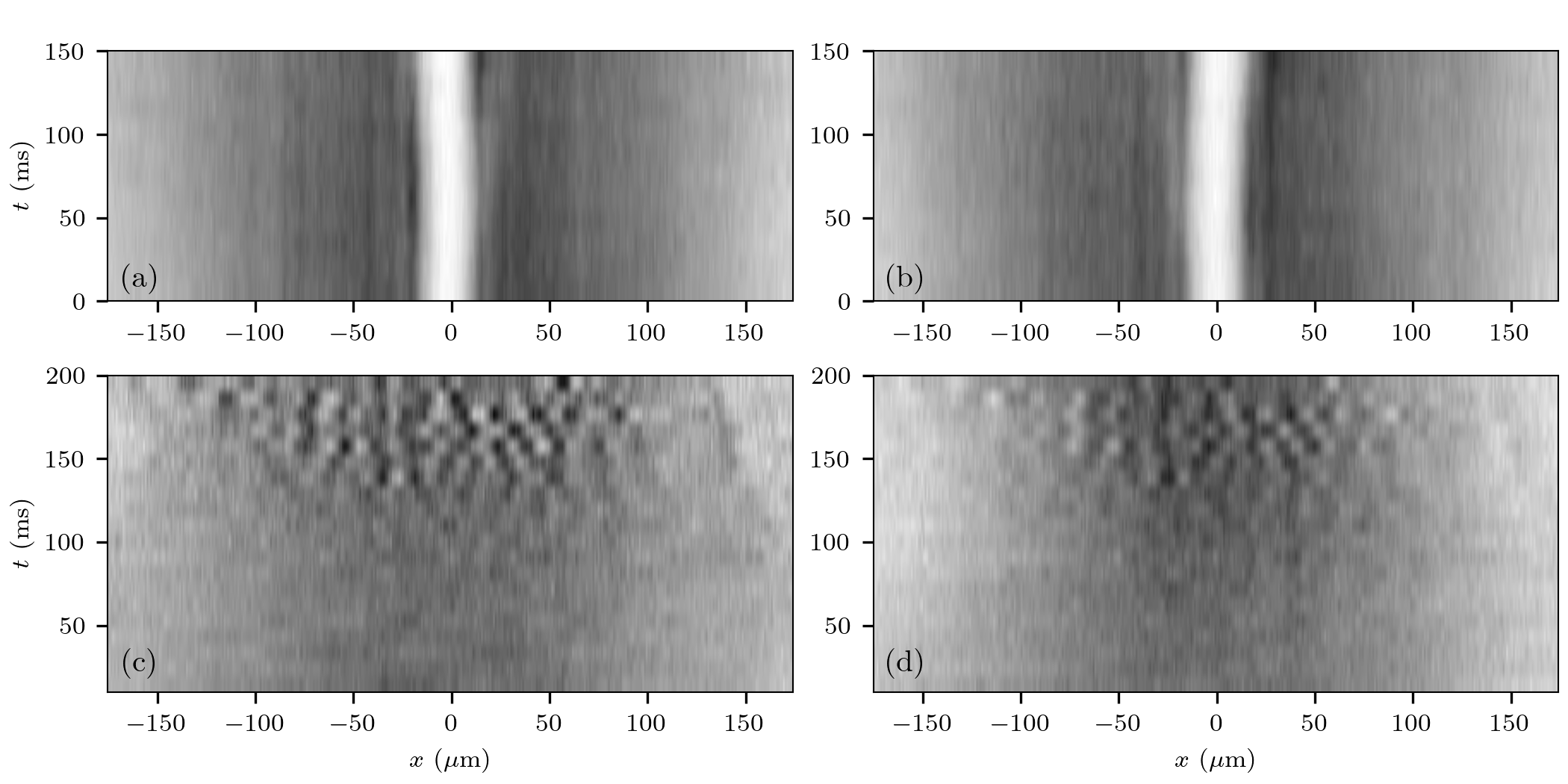}
    \caption{Results for control experiments. (a)-(b) An equal mixture of the $\lvert 1,-1\rangle$ and the $\lvert 1,0\rangle$ states (shown in panels (a) and (b), respectively), is prepared in the presence of an optical barrier, using a similar experimental procedure to the one described in the main text. 
    This specific mixture is miscible. 
    Each row is an average over 11 experimental realizations for evolution times from 0 to 150~ms in steps of 15 ms. MI generation is absent in the miscible interaction regime.
    (c)-(d) An equal mixture of the  $\lvert 2,0\rangle$ and the $\lvert 1,0\rangle$ states (shown in panels (c) and (d), respectively) is prepared without an optical barrier, using a similar experimental procedure to the one described in the main text.
    Each row is an average over 3 experimental realizations for evolution times from 0 to 200~ms in steps of 10~ms. 
    Notice that MI sets in spontaneously after approximately 125~ms across the central bulk of the condensate.
    In all cases above, each spin distribution is independently normalized such that darker colors represent higher relative atomic density per pixel.}
    \label{fig:controls}
\end{figure}

Additionally, in the one-component setting, one can also describe the envelope confining the modulations of the nonlinear stage of MI. This was found in Ref.~\cite{PRE2018v98p052220} to be
\begin{align}
\label{eq:envelope}
    |q_\pm(m)| = q_0 \pm \frac{q_0}{m} \left[2 - m - 2 (1 - m)  \frac{K(m)}{E(m)}\right] \,,
\end{align}
where $m$ is the elliptic parameter and $K(m)$ and $E(m)$ are the complete elliptic integrals of the first and second kind, respectively~\cite{NIST}.
The spatial mapping $m\mapsto x$ is given by $x/t = \sqrt{-g}\,\xi(m)$, with
\begin{align}
\xi(m) = \frac{4 q_0 \sqrt{1 - m}}{m E}  \frac{(2 - m) E  K + E^2 - 3 (1 - m) K^2}{\sqrt{E^2 - (1 - m)  K^2}} \,.
\label{e:xi}
\end{align}
Taking the limit of $m\to 0^+$ in Eq.~\eqref{e:xi} one can again obtain the speed $V_\textnormal{nls}$~\cite{PRE2018v98p052220,GinoDion} described by Eq.~(5) in the main text. 
Finally, as argued in the main text, Eq.~(\ref{eq:envelope}) adequately captures the MI wedge even in the weakly immiscible two-component repulsively interacting setting. 

Incidentally, we point out that a DSW is a self-similar expanding structure, and as such it does not have a characteristic wavelength.
Rather, its internal structure encompasses a range of length scales,
as has been discussed in various review works, e.g., see
\cite{EL201611}.

\section{Control cases: immiscibility and the barrier}

To further elucidate the importance of the various parts of the experimental setup, we next look specifically at the key role played by two ``ingredients'': (i)~immiscibility and (ii)~the repulsive barrier. 
First, to clarify the role of immiscible interactions for MI nucleation, we experimentally reproduce the procedure described in the main text but now using a miscible mixture instead. 
For the miscible mixture utilized, the interaction parameters do not trigger the emergence of MI. 
Specifically, starting with all atoms in the $\ket{1} \equiv \lvert 1,-1\rangle$ state, we use a radio frequency pulse to prepare an equal mixture of the $\ket{1} \equiv \lvert 1,-1\rangle$ and the $ \ket{2} \equiv \lvert 1,0\rangle$ states in the presence of the same optical barrier used in the main text. 
These hyperfine states are characterized by 3D $s$-wave scattering lengths $a_{11}=100.4a_0$, $a_{22}=100.86a_0$, and $a_{12}=a_{21}=100.41a_0$, with $a_0$ denoting the Bohr radius. 
This is a miscible combination, as indicated by the fact that $a_{22}-a_{12}^2/a_{11}>0$. 
Fig.~\ref{fig:controls}(a)-(b) presents the density dynamics of the $\lvert 1,-1\rangle$ state and $\lvert 1,0\rangle$ state, respectively. 
It can be readily seen that, in sharp contrast to the magnetically insensitive states used in the main body of this work, this mixture is solely subject to small drifts due to minute residual magnetic gradients along the long axis of the BEC.
As such, only trivial dynamics arise over the experimentally observed timescales, demonstrating the crucial role of immiscibility, and effectively attractive interactions, for MI generation. 
These dynamics have been independently verified using 3D GPE simulations following the experimental protocol. 
No MI was found even for longer timescales, which we have checked up to 500~ms. 

Next, to illustrate the crucial role of the presence of the repulsive barrier,
we consider the behavior of MI  without the barrier, which more closely models a uniform mixture.
Figure~\ref{fig:controls}(c)-(d) shows an initially equal mixture of the $\lvert 2,0\rangle$ and the $\lvert 1,0\rangle$ states, respectively. 
We observe that without the barrier to catalyze MI, no significant dynamics occur until approximately 125~ms have passed. 
At this point spontaneous MI sets in across the bulk of the condensate, first near the center of the condensate where the density is highest, in agreement with the background behavior for the barrier induced MI demonstrated in the main text.

\section{Dam-break problem and the Peregrine soliton}
\label{app:peregrine}

\begin{figure}[t!]
\centering
\includegraphics[width=\textwidth]{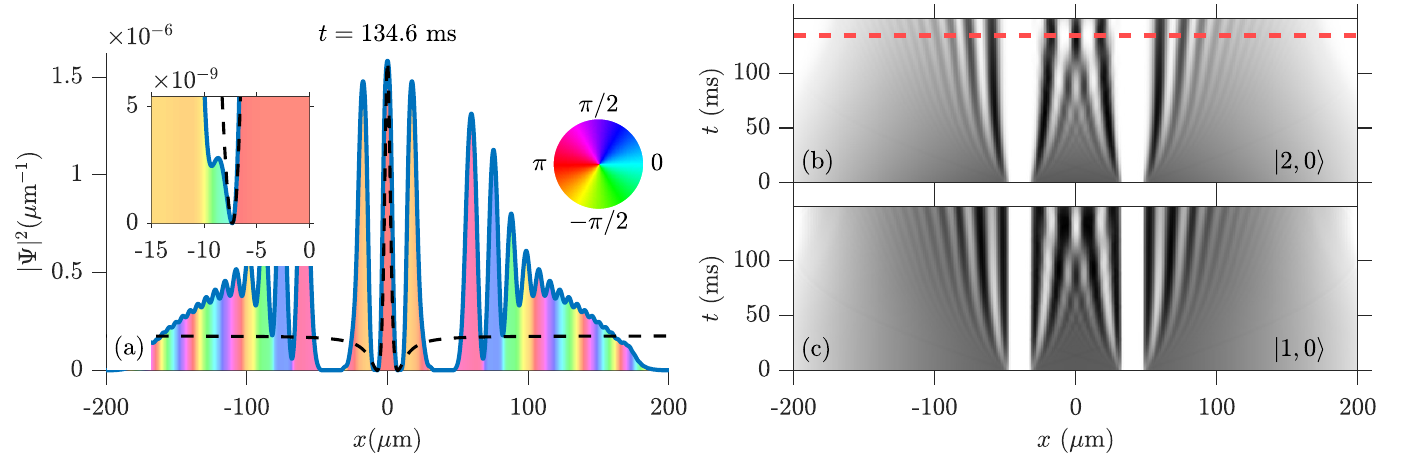}
\caption{
1D GPE simulations of the experimental procedure shown in Fig.~3. 
(a) Density profile of the $\ket{2,0}$ state at the instant at which a Peregrine-like structure forms around $x=0$.
In the inset, a magnification of the side dips is provided to clearly visualize the accompanied $\pi$-phase jump. 
The colormap denotes the phase at each spatial point.
The dashed black line refers to the analytical Peregrine solution [Eq.~(\ref{eq:peregrine})].       
(b)-(c) Spatiotemporal evolution of the $\ket{2,0}$ and $\ket{1,0}$ states, respectively in the presence of two repulsive potential wells.  
In panel (b), the red dotted line marks the time instant for the density profile shown in panel (a).}
\label{fig:peregrine}
\end{figure}

The Peregrine soliton is a solution of the focusing ($g<0$) NLS [Eq.~\eqref{eq:NLS}]  with non-vanishing boundary conditions.
In the context of the 1D GPE [Eq.~\eqref{eq:GPE_1D}] its equivalent solution has the form~\cite{peregrine1983water,Romero_theory}
\begin{equation}
P(x,t) = \sqrt{P_0}
\qty[1 - \frac{4 \qty(1 + 2 i \frac{t-t_0}{T_P})}{1+4 \qty(\frac{x-x_0}{L_P})^2 + 4\qty(\frac{t-t_0}{T_P})^2}] e^{i \frac{t-t_0}{T_P}},
\label{eq:peregrine}
\end{equation} 
where $T_P\omega_\perp=L_P^2/a_\perp^2=\hbar\omega_\perp/(|g| P_0)$. Here, $T_P$ and $L_P$ are the characteristic length and time scales of the Peregrine solution, respectively.
Also, $P_0$ represents the background density of a homogeneous system, and $g$ denotes the 1D single-component interaction strength.

In Fig.~3 of the main text, we discussed the experimentally monitored process of the interference of two  dam-break problems emanating from the presence of two symmetrically placed repulsive potential wells and leading to the spontaneous generation of a Peregrine-like structure around the center.
Figure~\ref{fig:peregrine} presents the corresponding dynamics obtained within 1D GPE numerical simulations emulating the aforementioned experimental procedure. 
The emergent spatiotemporal density evolution of the $\ket{2,0}$ and $\ket{1,0}$ states is depicted in Figs.~\ref{fig:peregrine}(b) and \ref{fig:peregrine}(c), respectively.
Note here the excellent agreement with the experimental results illustrated in Fig.~3 of the main text, where the generation of two counterpropagating DSWs triggered by the ensuing dam-break problems mimicked by the repulsive barriers is observed. 
Importantly, the interference of the DSWs within the region of the repulsive wells gives rise to the spontaneous nucleation of a Peregrine-like structure around $t_0 =134.6$~ms [see the red dotted line in panel (b)], 
as can be deduced by comparing to the fit of the Peregrine analytical waveform [Eq.~\eqref{eq:peregrine}] demonstrated by the dashed black line in panel (a). 
In the inset of panel (a) we also provide a magnified version of the dips of the density visualizing the characteristic $\pi$-phase jump accompanying a Peregrine waveform (see color wheel).

It is important to keep in mind that in our setup we are dealing with a two-component particle balanced immiscible mixture.
This lies away from the parametric regimes (high particle imbalance or attractive interactions) where a Peregrine soliton can be expected to be able to form, as it was the case of the experiment of Ref.~\cite{our2024PRL}.
Hence, in order to characterize the Peregrine-like structure shown in Fig.~3 and Fig.~\ref{fig:peregrine}, we used the experimentally accessible values $P_0=|\Psi_{\ket{2,0}}(x_0,t_0)|^2/9$ and $x_\pm=\pm 7.32 \ \mu$m, with $|\Psi_{\ket{2,0}}(x_\pm,t_0)|^2=0$.


\begin{thebibliography}{52}%
\makeatletter
\providecommand \@ifxundefined [1]{%
 \@ifx{#1\undefined}
}%
\providecommand \@ifnum [1]{%
 \ifnum #1\expandafter \@firstoftwo
 \else \expandafter \@secondoftwo
 \fi
}%
\providecommand \@ifx [1]{%
 \ifx #1\expandafter \@firstoftwo
 \else \expandafter \@secondoftwo
 \fi
}%
\providecommand \natexlab [1]{#1}%
\providecommand \enquote  [1]{``#1''}%
\providecommand \bibnamefont  [1]{#1}%
\providecommand \bibfnamefont [1]{#1}%
\providecommand \citenamefont [1]{#1}%
\providecommand \href@noop [0]{\@secondoftwo}%
\providecommand \href [0]{\begingroup \@sanitize@url \@href}%
\providecommand \@href[1]{\@@startlink{#1}\@@href}%
\providecommand \@@href[1]{\endgroup#1\@@endlink}%
\providecommand \@sanitize@url [0]{\catcode `\\12\catcode `\$12\catcode
  `\&12\catcode `\#12\catcode `\^12\catcode `\_12\catcode `\%12\relax}%
\providecommand \@@startlink[1]{}%
\providecommand \@@endlink[0]{}%
\providecommand \url  [0]{\begingroup\@sanitize@url \@url }%
\providecommand \@url [1]{\endgroup\@href {#1}{\urlprefix }}%
\providecommand \urlprefix  [0]{URL }%
\providecommand \Eprint [0]{\href }%
\providecommand \doibase [0]{https://doi.org/}%
\providecommand \selectlanguage [0]{\@gobble}%
\providecommand \bibinfo  [0]{\@secondoftwo}%
\providecommand \bibfield  [0]{\@secondoftwo}%
\providecommand \translation [1]{[#1]}%
\providecommand \BibitemOpen [0]{}%
\providecommand \bibitemStop [0]{}%
\providecommand \bibitemNoStop [0]{.\EOS\space}%
\providecommand \EOS [0]{\spacefactor3000\relax}%
\providecommand \BibitemShut  [1]{\csname bibitem#1\endcsname}%
\let\auto@bib@innerbib\@empty
%</preamble>
\bibitem [{\citenamefont {Zakharov}\ and\ \citenamefont
  {Ostrovsky}(2009)}]{PhysicaD238p540}%
  \BibitemOpen
  \bibfield  {author} {\bibinfo {author} {\bibfnamefont {V.}~\bibnamefont
  {Zakharov}}\ and\ \bibinfo {author} {\bibfnamefont {L.}~\bibnamefont
  {Ostrovsky}},\ }\href@noop {} {\bibfield  {journal} {\bibinfo  {journal}
  {Physica D}\ }\textbf {\bibinfo {volume} {238}},\ \bibinfo {pages} {540}
  (\bibinfo {year} {2009})}\BibitemShut {NoStop}%
\bibitem [{\citenamefont {Everitt}\ \emph
  {et~al.}(2017{\natexlab{a}})\citenamefont {Everitt}, \citenamefont
  {Sooriyabandara}, \citenamefont {Guasoni}, \citenamefont {Wigley},
  \citenamefont {Wei}, \citenamefont {McDonald}, \citenamefont {Hardman},
  \citenamefont {Manju}, \citenamefont {Close}, \citenamefont {Kuhn},
  \citenamefont {Szigeti}, \citenamefont {Kivshar},\ and\ \citenamefont
  {Robins}}]{PhysRevA.96.041601}%
  \BibitemOpen
  \bibfield  {author} {\bibinfo {author} {\bibfnamefont {P.~J.}\ \bibnamefont
  {Everitt}}, \bibinfo {author} {\bibfnamefont {M.~A.}\ \bibnamefont
  {Sooriyabandara}}, \bibinfo {author} {\bibfnamefont {M.}~\bibnamefont
  {Guasoni}}, \bibinfo {author} {\bibfnamefont {P.~B.}\ \bibnamefont {Wigley}},
  \bibinfo {author} {\bibfnamefont {C.~H.}\ \bibnamefont {Wei}}, \bibinfo
  {author} {\bibfnamefont {G.~D.}\ \bibnamefont {McDonald}}, \bibinfo {author}
  {\bibfnamefont {K.~S.}\ \bibnamefont {Hardman}}, \bibinfo {author}
  {\bibfnamefont {P.}~\bibnamefont {Manju}}, \bibinfo {author} {\bibfnamefont
  {J.~D.}\ \bibnamefont {Close}}, \bibinfo {author} {\bibfnamefont {C.~C.~N.}\
  \bibnamefont {Kuhn}}, \bibinfo {author} {\bibfnamefont {S.~S.}\ \bibnamefont
  {Szigeti}}, \bibinfo {author} {\bibfnamefont {Y.~S.}\ \bibnamefont
  {Kivshar}},\ and\ \bibinfo {author} {\bibfnamefont {N.~P.}\ \bibnamefont
  {Robins}},\ }\href {https://doi.org/10.1103/PhysRevA.96.041601} {\bibfield
  {journal} {\bibinfo  {journal} {Phys. Rev. A}\ }\textbf {\bibinfo {volume}
  {96}},\ \bibinfo {pages} {041601} (\bibinfo {year}
  {2017}{\natexlab{a}})}\BibitemShut {NoStop}%
\bibitem [{\citenamefont {Chen}\ and\ \citenamefont
  {Hung}(2020{\natexlab{a}})}]{PhysRevLett.125.250401}%
  \BibitemOpen
  \bibfield  {author} {\bibinfo {author} {\bibfnamefont {C.-A.}\ \bibnamefont
  {Chen}}\ and\ \bibinfo {author} {\bibfnamefont {C.-L.}\ \bibnamefont
  {Hung}},\ }\href {https://doi.org/10.1103/PhysRevLett.125.250401} {\bibfield
  {journal} {\bibinfo  {journal} {Phys. Rev. Lett.}\ }\textbf {\bibinfo
  {volume} {125}},\ \bibinfo {pages} {250401} (\bibinfo {year}
  {2020}{\natexlab{a}})}\BibitemShut {NoStop}%
\bibitem [{\citenamefont {Kevrekidis}\ \emph {et~al.}(2015)\citenamefont
  {Kevrekidis}, \citenamefont {Frantzeskakis},\ and\ \citenamefont
  {{Carretero-Gonz{\'a}lez}}}]{Kevrekidis2015}%
  \BibitemOpen
  \bibfield  {author} {\bibinfo {author} {\bibfnamefont {P.~G.}\ \bibnamefont
  {Kevrekidis}}, \bibinfo {author} {\bibfnamefont {D.~J.}\ \bibnamefont
  {Frantzeskakis}},\ and\ \bibinfo {author} {\bibfnamefont {R.}~\bibnamefont
  {{Carretero-Gonz{\'a}lez}}},\ }\href
  {https://doi.org/10.1137/1.9781611973945} {\emph {\bibinfo {title} {The
  {{Defocusing Nonlinear Schr\"odinger Equation}}: From Dark Soliton to
  Vortices and Vortex Rings}}},\ Other {{Titles}} in {{Applied Mathematics}}\
  (\bibinfo  {publisher} {{Society for Industrial and Applied Mathematics}},\
  \bibinfo {address} {{Philadelphia}},\ \bibinfo {year} {2015})\BibitemShut
  {NoStop}%
\bibitem [{\citenamefont {Kivshar}\ and\ \citenamefont
  {Agrawal}(2003)}]{Kivshar2003}%
  \BibitemOpen
  \bibfield  {author} {\bibinfo {author} {\bibfnamefont {Y.~S.}\ \bibnamefont
  {Kivshar}}\ and\ \bibinfo {author} {\bibfnamefont {G.~P.}\ \bibnamefont
  {Agrawal}},\ }\href {https://doi.org/10.1016/B978-0-12-410590-4.X5000-1}
  {\emph {\bibinfo {title} {Optical Solitons: From Fibers to Photonic
  Crystals}}}\ (\bibinfo  {publisher} {Academic Press},\ \bibinfo {year}
  {2003})\ pp.\ \bibinfo {pages} {1--540}\BibitemShut {NoStop}%
\bibitem [{\citenamefont {Ablowitz}(2011)}]{ablowitz2}%
  \BibitemOpen
  \bibfield  {author} {\bibinfo {author} {\bibfnamefont {M.}~\bibnamefont
  {Ablowitz}},\ }\href {https://doi.org/10.1017/cbo9780511998324} {\emph
  {\bibinfo {title} {Nonlinear Dispersive Waves, Asymptotic Analysis and
  Solitons}}}\ (\bibinfo  {publisher} {Cambridge University Press, Cambridge},\
  \bibinfo {year} {2011})\BibitemShut {NoStop}%
\bibitem [{\citenamefont {Kono}\ and\ \citenamefont {Skori{\'c}}(2010)}]{kono}%
  \BibitemOpen
  \bibfield  {author} {\bibinfo {author} {\bibfnamefont {M.}~\bibnamefont
  {Kono}}\ and\ \bibinfo {author} {\bibfnamefont {M.}~\bibnamefont
  {Skori{\'c}}},\ }\href@noop {} {\emph {\bibinfo {title} {Nonlinear Physics of
  Plasmas}}}\ (\bibinfo  {publisher} {Springer-Verlag, Heidelberg},\ \bibinfo
  {year} {2010})\BibitemShut {NoStop}%
\bibitem [{\citenamefont {Zakharov}\ and\ \citenamefont
  {Gelash}(2013)}]{ZakhPRL}%
  \BibitemOpen
  \bibfield  {author} {\bibinfo {author} {\bibfnamefont {V.~E.}\ \bibnamefont
  {Zakharov}}\ and\ \bibinfo {author} {\bibfnamefont {A.~A.}\ \bibnamefont
  {Gelash}},\ }\href {https://doi.org/10.1103/PhysRevLett.111.054101}
  {\bibfield  {journal} {\bibinfo  {journal} {Phys. Rev. Lett.}\ }\textbf
  {\bibinfo {volume} {111}},\ \bibinfo {pages} {054101} (\bibinfo {year}
  {2013})}\BibitemShut {NoStop}%
\bibitem [{\citenamefont {Biondini}\ and\ \citenamefont
  {Mantzavinos}(2016)}]{GinoDion}%
  \BibitemOpen
  \bibfield  {author} {\bibinfo {author} {\bibfnamefont {G.}~\bibnamefont
  {Biondini}}\ and\ \bibinfo {author} {\bibfnamefont {D.}~\bibnamefont
  {Mantzavinos}},\ }\href {https://doi.org/10.1103/PhysRevLett.116.043902}
  {\bibfield  {journal} {\bibinfo  {journal} {Phys. Rev. Lett.}\ }\textbf
  {\bibinfo {volume} {116}},\ \bibinfo {pages} {043902} (\bibinfo {year}
  {2016})}\BibitemShut {NoStop}%
\bibitem [{\citenamefont {El'}\ \emph {et~al.}(1993)\citenamefont {El'},
  \citenamefont {Gurevich}, \citenamefont {Khodorovskiǐ},\ and\ \citenamefont
  {Krylov}}]{EL1993357}%
  \BibitemOpen
  \bibfield  {author} {\bibinfo {author} {\bibfnamefont {G.}~\bibnamefont
  {El'}}, \bibinfo {author} {\bibfnamefont {A.}~\bibnamefont {Gurevich}},
  \bibinfo {author} {\bibfnamefont {V.}~\bibnamefont {Khodorovskiǐ}},\ and\
  \bibinfo {author} {\bibfnamefont {A.}~\bibnamefont {Krylov}},\ }\href
  {https://doi.org/https://doi.org/10.1016/0375-9601(93)90015-R} {\bibfield
  {journal} {\bibinfo  {journal} {Physics Letters A}\ }\textbf {\bibinfo
  {volume} {177}},\ \bibinfo {pages} {357} (\bibinfo {year}
  {1993})}\BibitemShut {NoStop}%
\bibitem [{\citenamefont {Whitham}(1965)}]{whitham1965non}%
  \BibitemOpen
  \bibfield  {author} {\bibinfo {author} {\bibfnamefont {G.~B.}\ \bibnamefont
  {Whitham}},\ }\href@noop {} {\bibfield  {journal} {\bibinfo  {journal}
  {Proceedings of the Royal Society of London. Series A. Mathematical and
  Physical Sciences}\ }\textbf {\bibinfo {volume} {283}},\ \bibinfo {pages}
  {238} (\bibinfo {year} {1965})}\BibitemShut {NoStop}%
\bibitem [{\citenamefont {Biondini}\ and\ \citenamefont
  {Kova\v{c}i\v{c}}(2014)}]{JMP55p031506}%
  \BibitemOpen
  \bibfield  {author} {\bibinfo {author} {\bibfnamefont {G.}~\bibnamefont
  {Biondini}}\ and\ \bibinfo {author} {\bibfnamefont {G.}~\bibnamefont
  {Kova\v{c}i\v{c}}},\ }\href@noop {} {\bibfield  {journal} {\bibinfo
  {journal} {J. Math. Phys.}\ }\textbf {\bibinfo {volume} {55}},\ \bibinfo
  {pages} {031506} (\bibinfo {year} {2014})}\BibitemShut {NoStop}%
\bibitem [{\citenamefont {Kraych}\ \emph {et~al.}(2019)\citenamefont {Kraych},
  \citenamefont {Suret}, \citenamefont {El},\ and\ \citenamefont
  {Randoux}}]{Kraych2019}%
  \BibitemOpen
  \bibfield  {author} {\bibinfo {author} {\bibfnamefont {A.~E.}\ \bibnamefont
  {Kraych}}, \bibinfo {author} {\bibfnamefont {P.}~\bibnamefont {Suret}},
  \bibinfo {author} {\bibfnamefont {G.}~\bibnamefont {El}},\ and\ \bibinfo
  {author} {\bibfnamefont {S.}~\bibnamefont {Randoux}},\ }\href
  {https://doi.org/10.1103/physrevlett.122.054101} {\bibfield  {journal}
  {\bibinfo  {journal} {Physical Review Letters}\ }\textbf {\bibinfo {volume}
  {122}},\ \bibinfo {pages} {054101} (\bibinfo {year} {2019})}\BibitemShut
  {NoStop}%
\bibitem [{\citenamefont {Copie}\ \emph {et~al.}(2020)\citenamefont {Copie},
  \citenamefont {Randoux},\ and\ \citenamefont {Suret}}]{COPIE2020100037}%
  \BibitemOpen
  \bibfield  {author} {\bibinfo {author} {\bibfnamefont {F.}~\bibnamefont
  {Copie}}, \bibinfo {author} {\bibfnamefont {S.}~\bibnamefont {Randoux}},\
  and\ \bibinfo {author} {\bibfnamefont {P.}~\bibnamefont {Suret}},\ }\href
  {https://doi.org/https://doi.org/10.1016/j.revip.2019.100037} {\bibfield
  {journal} {\bibinfo  {journal} {Reviews in Physics}\ }\textbf {\bibinfo
  {volume} {5}},\ \bibinfo {pages} {100037} (\bibinfo {year}
  {2020})}\BibitemShut {NoStop}%
\bibitem [{\citenamefont {Bonnefoy}\ \emph {et~al.}(2020)\citenamefont
  {Bonnefoy}, \citenamefont {Tikan}, \citenamefont {Copie}, \citenamefont
  {Suret}, \citenamefont {Ducrozet}, \citenamefont {Prabhudesai}, \citenamefont
  {Michel}, \citenamefont {Cazaubiel}, \citenamefont {Falcon}, \citenamefont
  {El},\ and\ \citenamefont {Randoux}}]{Suret1}%
  \BibitemOpen
  \bibfield  {author} {\bibinfo {author} {\bibfnamefont {F.}~\bibnamefont
  {Bonnefoy}}, \bibinfo {author} {\bibfnamefont {A.}~\bibnamefont {Tikan}},
  \bibinfo {author} {\bibfnamefont {F.}~\bibnamefont {Copie}}, \bibinfo
  {author} {\bibfnamefont {P.}~\bibnamefont {Suret}}, \bibinfo {author}
  {\bibfnamefont {G.}~\bibnamefont {Ducrozet}}, \bibinfo {author}
  {\bibfnamefont {G.}~\bibnamefont {Prabhudesai}}, \bibinfo {author}
  {\bibfnamefont {G.}~\bibnamefont {Michel}}, \bibinfo {author} {\bibfnamefont
  {A.}~\bibnamefont {Cazaubiel}}, \bibinfo {author} {\bibfnamefont
  {E.}~\bibnamefont {Falcon}}, \bibinfo {author} {\bibfnamefont
  {G.}~\bibnamefont {El}},\ and\ \bibinfo {author} {\bibfnamefont
  {S.}~\bibnamefont {Randoux}},\ }\href
  {https://doi.org/10.1103/PhysRevFluids.5.034802} {\bibfield  {journal}
  {\bibinfo  {journal} {Phys. Rev. Fluids}\ }\textbf {\bibinfo {volume} {5}},\
  \bibinfo {pages} {034802} (\bibinfo {year} {2020})}\BibitemShut {NoStop}%
\bibitem [{\citenamefont {Biondini}\ \emph {et~al.}(2018)\citenamefont
  {Biondini}, \citenamefont {Li}, \citenamefont {Mantzavinos},\ and\
  \citenamefont {Trillo}}]{SIREV2018v60p888}%
  \BibitemOpen
  \bibfield  {author} {\bibinfo {author} {\bibfnamefont {G.}~\bibnamefont
  {Biondini}}, \bibinfo {author} {\bibfnamefont {S.}~\bibnamefont {Li}},
  \bibinfo {author} {\bibfnamefont {D.}~\bibnamefont {Mantzavinos}},\ and\
  \bibinfo {author} {\bibfnamefont {S.}~\bibnamefont {Trillo}},\ }\href
  {http://www.math.buffalo.edu/~biondini/papers/sirev2018v60p888.pdf}
  {\bibfield  {journal} {\bibinfo  {journal} {SIAM Review}\ }\textbf {\bibinfo
  {volume} {60}},\ \bibinfo {pages} {888} (\bibinfo {year} {2018})}\BibitemShut
  {NoStop}%
\bibitem [{\citenamefont {Bloch}\ \emph {et~al.}(2008)\citenamefont {Bloch},
  \citenamefont {Dalibard},\ and\ \citenamefont {Zwerger}}]{Bloch_many_body}%
  \BibitemOpen
  \bibfield  {author} {\bibinfo {author} {\bibfnamefont {I.}~\bibnamefont
  {Bloch}}, \bibinfo {author} {\bibfnamefont {J.}~\bibnamefont {Dalibard}},\
  and\ \bibinfo {author} {\bibfnamefont {W.}~\bibnamefont {Zwerger}},\ }\href
  {https://doi.org/10.1103/RevModPhys.80.885} {\bibfield  {journal} {\bibinfo
  {journal} {Rev. Mod. Phys.}\ }\textbf {\bibinfo {volume} {80}},\ \bibinfo
  {pages} {885} (\bibinfo {year} {2008})}\BibitemShut {NoStop}%
\bibitem [{\citenamefont {Gross}\ and\ \citenamefont
  {Bloch}(2017)}]{gross2017quantum}%
  \BibitemOpen
  \bibfield  {author} {\bibinfo {author} {\bibfnamefont {C.}~\bibnamefont
  {Gross}}\ and\ \bibinfo {author} {\bibfnamefont {I.}~\bibnamefont {Bloch}},\
  }\href@noop {} {\bibfield  {journal} {\bibinfo  {journal} {Science}\ }\textbf
  {\bibinfo {volume} {357}},\ \bibinfo {pages} {995} (\bibinfo {year}
  {2017})}\BibitemShut {NoStop}%
\bibitem [{\citenamefont {Mistakidis}\ \emph {et~al.}(2023)\citenamefont
  {Mistakidis}, \citenamefont {Volosniev}, \citenamefont {Barfknecht},
  \citenamefont {Fogarty}, \citenamefont {Busch}, \citenamefont {Foerster},
  \citenamefont {Schmelcher},\ and\ \citenamefont
  {Zinner}}]{Mistakidis_review}%
  \BibitemOpen
  \bibfield  {author} {\bibinfo {author} {\bibfnamefont {S.~I.}\ \bibnamefont
  {Mistakidis}}, \bibinfo {author} {\bibfnamefont {A.~G.}\ \bibnamefont
  {Volosniev}}, \bibinfo {author} {\bibfnamefont {R.~E.}\ \bibnamefont
  {Barfknecht}}, \bibinfo {author} {\bibfnamefont {T.}~\bibnamefont {Fogarty}},
  \bibinfo {author} {\bibfnamefont {T.}~\bibnamefont {Busch}}, \bibinfo
  {author} {\bibfnamefont {A.}~\bibnamefont {Foerster}}, \bibinfo {author}
  {\bibfnamefont {P.}~\bibnamefont {Schmelcher}},\ and\ \bibinfo {author}
  {\bibfnamefont {N.~T.}\ \bibnamefont {Zinner}},\ }\href
  {https://doi.org/https://doi.org/10.1016/j.physrep.2023.10.004} {\bibfield
  {journal} {\bibinfo  {journal} {Physics Reports}\ }\textbf {\bibinfo {volume}
  {1042}},\ \bibinfo {pages} {1} (\bibinfo {year} {2023})}\BibitemShut
  {NoStop}%
\bibitem [{\citenamefont {Strecker}\ \emph {et~al.}(2002)\citenamefont
  {Strecker}, \citenamefont {Partridge}, \citenamefont {Truscott},\ and\
  \citenamefont {Hulet}}]{Strecker2002}%
  \BibitemOpen
  \bibfield  {author} {\bibinfo {author} {\bibfnamefont {K.~E.}\ \bibnamefont
  {Strecker}}, \bibinfo {author} {\bibfnamefont {G.~B.}\ \bibnamefont
  {Partridge}}, \bibinfo {author} {\bibfnamefont {A.~G.}\ \bibnamefont
  {Truscott}},\ and\ \bibinfo {author} {\bibfnamefont {R.~G.}\ \bibnamefont
  {Hulet}},\ }\href {https://doi.org/10.1038/nature747} {\bibfield  {journal}
  {\bibinfo  {journal} {Nature}\ }\textbf {\bibinfo {volume} {417}},\ \bibinfo
  {pages} {150} (\bibinfo {year} {2002})}\BibitemShut {NoStop}%
\bibitem [{\citenamefont {Strecker}\ \emph {et~al.}(2003)\citenamefont
  {Strecker}, \citenamefont {Partridge}, \citenamefont {Truscott},\ and\
  \citenamefont {Hulet}}]{Strecker2003}%
  \BibitemOpen
  \bibfield  {author} {\bibinfo {author} {\bibfnamefont {K.~E.}\ \bibnamefont
  {Strecker}}, \bibinfo {author} {\bibfnamefont {G.~B.}\ \bibnamefont
  {Partridge}}, \bibinfo {author} {\bibfnamefont {A.~G.}\ \bibnamefont
  {Truscott}},\ and\ \bibinfo {author} {\bibfnamefont {R.~G.}\ \bibnamefont
  {Hulet}},\ }\href {https://doi.org/10.1088/1367-2630/5/1/373} {\bibfield
  {journal} {\bibinfo  {journal} {New J. Phys.}\ }\textbf {\bibinfo {volume}
  {5}},\ \bibinfo {pages} {73} (\bibinfo {year} {2003})}\BibitemShut {NoStop}%
\bibitem [{\citenamefont {Nguyen}\ \emph {et~al.}(2017)\citenamefont {Nguyen},
  \citenamefont {Luo},\ and\ \citenamefont {Hulet}}]{nguyen}%
  \BibitemOpen
  \bibfield  {author} {\bibinfo {author} {\bibfnamefont {J.~H.~V.}\
  \bibnamefont {Nguyen}}, \bibinfo {author} {\bibfnamefont {D.}~\bibnamefont
  {Luo}},\ and\ \bibinfo {author} {\bibfnamefont {R.~G.}\ \bibnamefont
  {Hulet}},\ }\href {https://doi.org/10.1126/science.aal3220} {\bibfield
  {journal} {\bibinfo  {journal} {Science}\ }\textbf {\bibinfo {volume}
  {356}},\ \bibinfo {pages} {422} (\bibinfo {year} {2017})}\BibitemShut
  {NoStop}%
\bibitem [{\citenamefont {Everitt}\ \emph
  {et~al.}(2017{\natexlab{b}})\citenamefont {Everitt}, \citenamefont
  {Sooriyabandara}, \citenamefont {Guasoni}, \citenamefont {Wigley},
  \citenamefont {Wei}, \citenamefont {McDonald}, \citenamefont {Hardman},
  \citenamefont {Manju}, \citenamefont {Close}, \citenamefont {Kuhn},
  \citenamefont {Szigeti}, \citenamefont {Kivshar},\ and\ \citenamefont
  {Robins}}]{everitt}%
  \BibitemOpen
  \bibfield  {author} {\bibinfo {author} {\bibfnamefont {P.~J.}\ \bibnamefont
  {Everitt}}, \bibinfo {author} {\bibfnamefont {M.~A.}\ \bibnamefont
  {Sooriyabandara}}, \bibinfo {author} {\bibfnamefont {M.}~\bibnamefont
  {Guasoni}}, \bibinfo {author} {\bibfnamefont {P.~B.}\ \bibnamefont {Wigley}},
  \bibinfo {author} {\bibfnamefont {C.~H.}\ \bibnamefont {Wei}}, \bibinfo
  {author} {\bibfnamefont {G.~D.}\ \bibnamefont {McDonald}}, \bibinfo {author}
  {\bibfnamefont {K.~S.}\ \bibnamefont {Hardman}}, \bibinfo {author}
  {\bibfnamefont {P.}~\bibnamefont {Manju}}, \bibinfo {author} {\bibfnamefont
  {J.~D.}\ \bibnamefont {Close}}, \bibinfo {author} {\bibfnamefont {C.~C.~N.}\
  \bibnamefont {Kuhn}}, \bibinfo {author} {\bibfnamefont {S.~S.}\ \bibnamefont
  {Szigeti}}, \bibinfo {author} {\bibfnamefont {Y.~S.}\ \bibnamefont
  {Kivshar}},\ and\ \bibinfo {author} {\bibfnamefont {N.~P.}\ \bibnamefont
  {Robins}},\ }\href {https://doi.org/10.1103/PhysRevA.96.041601} {\bibfield
  {journal} {\bibinfo  {journal} {Phys. Rev. A}\ }\textbf {\bibinfo {volume}
  {96}},\ \bibinfo {pages} {041601} (\bibinfo {year}
  {2017}{\natexlab{b}})}\BibitemShut {NoStop}%
\bibitem [{\citenamefont {Chen}\ and\ \citenamefont
  {Hung}(2020{\natexlab{b}})}]{Lung_Townes2D}%
  \BibitemOpen
  \bibfield  {author} {\bibinfo {author} {\bibfnamefont {C.-A.}\ \bibnamefont
  {Chen}}\ and\ \bibinfo {author} {\bibfnamefont {C.-L.}\ \bibnamefont
  {Hung}},\ }\href {https://doi.org/10.1103/PhysRevLett.125.250401} {\bibfield
  {journal} {\bibinfo  {journal} {Phys. Rev. Lett.}\ }\textbf {\bibinfo
  {volume} {125}},\ \bibinfo {pages} {250401} (\bibinfo {year}
  {2020}{\natexlab{b}})}\BibitemShut {NoStop}%
\bibitem [{\citenamefont {Chen}\ and\ \citenamefont
  {Hung}(2021)}]{Lung_Townes}%
  \BibitemOpen
  \bibfield  {author} {\bibinfo {author} {\bibfnamefont {C.-A.}\ \bibnamefont
  {Chen}}\ and\ \bibinfo {author} {\bibfnamefont {C.-L.}\ \bibnamefont
  {Hung}},\ }\href {https://doi.org/10.1103/PhysRevLett.127.023604} {\bibfield
  {journal} {\bibinfo  {journal} {Phys. Rev. Lett.}\ }\textbf {\bibinfo
  {volume} {127}},\ \bibinfo {pages} {023604} (\bibinfo {year}
  {2021})}\BibitemShut {NoStop}%
\bibitem [{\citenamefont {Banerjee}\ \emph {et~al.}(2024)\citenamefont
  {Banerjee}, \citenamefont {Zhou}, \citenamefont {Tiwari}, \citenamefont
  {Tamura}, \citenamefont {Li}, \citenamefont {Kevrekidis}, \citenamefont
  {Mistakidis}, \citenamefont {Walther},\ and\ \citenamefont
  {Hung}}]{banerjee2024collapse}%
  \BibitemOpen
  \bibfield  {author} {\bibinfo {author} {\bibfnamefont {S.}~\bibnamefont
  {Banerjee}}, \bibinfo {author} {\bibfnamefont {K.}~\bibnamefont {Zhou}},
  \bibinfo {author} {\bibfnamefont {S.~K.}\ \bibnamefont {Tiwari}}, \bibinfo
  {author} {\bibfnamefont {H.}~\bibnamefont {Tamura}}, \bibinfo {author}
  {\bibfnamefont {R.}~\bibnamefont {Li}}, \bibinfo {author} {\bibfnamefont
  {P.}~\bibnamefont {Kevrekidis}}, \bibinfo {author} {\bibfnamefont {S.~I.}\
  \bibnamefont {Mistakidis}}, \bibinfo {author} {\bibfnamefont
  {V.}~\bibnamefont {Walther}},\ and\ \bibinfo {author} {\bibfnamefont {C.-L.}\
  \bibnamefont {Hung}},\ }\href@noop {} {\bibfield  {journal} {\bibinfo
  {journal} {arXiv:2406.00863}\ } (\bibinfo {year} {2024})}\BibitemShut
  {NoStop}%
\bibitem [{\citenamefont {Kartashov}\ \emph {et~al.}(2011)\citenamefont
  {Kartashov}, \citenamefont {Malomed},\ and\ \citenamefont
  {Torner}}]{Kartashov_review}%
  \BibitemOpen
  \bibfield  {author} {\bibinfo {author} {\bibfnamefont {Y.~V.}\ \bibnamefont
  {Kartashov}}, \bibinfo {author} {\bibfnamefont {B.~A.}\ \bibnamefont
  {Malomed}},\ and\ \bibinfo {author} {\bibfnamefont {L.}~\bibnamefont
  {Torner}},\ }\href {https://doi.org/10.1103/RevModPhys.83.247} {\bibfield
  {journal} {\bibinfo  {journal} {Rev. Mod. Phys.}\ }\textbf {\bibinfo {volume}
  {83}},\ \bibinfo {pages} {247} (\bibinfo {year} {2011})}\BibitemShut
  {NoStop}%
\bibitem [{\citenamefont {{Bakkali-Hassani}}\ \emph {et~al.}(2021)\citenamefont
  {{Bakkali-Hassani}}, \citenamefont {Maury}, \citenamefont {Zou},
  \citenamefont {Le~Cerf}, \citenamefont {{Saint-Jalm}}, \citenamefont
  {Castilho}, \citenamefont {Nascimbene}, \citenamefont {Dalibard},\ and\
  \citenamefont {Beugnon}}]{Bakkali-Hassani2021}%
  \BibitemOpen
  \bibfield  {author} {\bibinfo {author} {\bibfnamefont {B.}~\bibnamefont
  {{Bakkali-Hassani}}}, \bibinfo {author} {\bibfnamefont {C.}~\bibnamefont
  {Maury}}, \bibinfo {author} {\bibfnamefont {Y.-Q.}\ \bibnamefont {Zou}},
  \bibinfo {author} {\bibfnamefont {{\'E}.}~\bibnamefont {Le~Cerf}}, \bibinfo
  {author} {\bibfnamefont {R.}~\bibnamefont {{Saint-Jalm}}}, \bibinfo {author}
  {\bibfnamefont {P.~C.~M.}\ \bibnamefont {Castilho}}, \bibinfo {author}
  {\bibfnamefont {S.}~\bibnamefont {Nascimbene}}, \bibinfo {author}
  {\bibfnamefont {J.}~\bibnamefont {Dalibard}},\ and\ \bibinfo {author}
  {\bibfnamefont {J.}~\bibnamefont {Beugnon}},\ }\href
  {https://doi.org/10.1103/PhysRevLett.127.023603} {\bibfield  {journal}
  {\bibinfo  {journal} {Phys. Rev. Lett.}\ }\textbf {\bibinfo {volume} {127}},\
  \bibinfo {pages} {023603} (\bibinfo {year} {2021})}\BibitemShut {NoStop}%
\bibitem [{\citenamefont {Romero-Ros}\ \emph {et~al.}(2024)\citenamefont
  {Romero-Ros}, \citenamefont {Katsimiga}, \citenamefont {Mistakidis},
  \citenamefont {Mossman}, \citenamefont {Biondini}, \citenamefont
  {Schmelcher}, \citenamefont {Engels},\ and\ \citenamefont
  {Kevrekidis}}]{our2024PRL}%
  \BibitemOpen
  \bibfield  {author} {\bibinfo {author} {\bibfnamefont {A.}~\bibnamefont
  {Romero-Ros}}, \bibinfo {author} {\bibfnamefont {G.~C.}\ \bibnamefont
  {Katsimiga}}, \bibinfo {author} {\bibfnamefont {S.~I.}\ \bibnamefont
  {Mistakidis}}, \bibinfo {author} {\bibfnamefont {S.}~\bibnamefont {Mossman}},
  \bibinfo {author} {\bibfnamefont {G.}~\bibnamefont {Biondini}}, \bibinfo
  {author} {\bibfnamefont {P.}~\bibnamefont {Schmelcher}}, \bibinfo {author}
  {\bibfnamefont {P.}~\bibnamefont {Engels}},\ and\ \bibinfo {author}
  {\bibfnamefont {P.~G.}\ \bibnamefont {Kevrekidis}},\ }\href
  {https://doi.org/10.1103/PhysRevLett.132.033402} {\bibfield  {journal}
  {\bibinfo  {journal} {Phys. Rev. Lett.}\ }\textbf {\bibinfo {volume} {132}},\
  \bibinfo {pages} {033402} (\bibinfo {year} {2024})}\BibitemShut {NoStop}%
\bibitem [{\citenamefont {Whitham}(2011)}]{whitham2011linear}%
  \BibitemOpen
  \bibfield  {author} {\bibinfo {author} {\bibfnamefont {G.~B.}\ \bibnamefont
  {Whitham}},\ }\href@noop {} {\emph {\bibinfo {title} {Linear and nonlinear
  waves}}}\ (\bibinfo  {publisher} {John Wiley \& Sons},\ \bibinfo {year}
  {2011})\BibitemShut {NoStop}%
\bibitem [{\citenamefont {El}\ and\ \citenamefont {Hoefer}(2016)}]{EL201611}%
  \BibitemOpen
  \bibfield  {author} {\bibinfo {author} {\bibfnamefont {G.}~\bibnamefont
  {El}}\ and\ \bibinfo {author} {\bibfnamefont {M.}~\bibnamefont {Hoefer}},\
  }\href {https://doi.org/https://doi.org/10.1016/j.physd.2016.04.006}
  {\bibfield  {journal} {\bibinfo  {journal} {Phys. D}\ }\textbf {\bibinfo
  {volume} {333}},\ \bibinfo {pages} {11} (\bibinfo {year} {2016})}\BibitemShut
  {NoStop}%
\bibitem [{\citenamefont {Dutton}\ and\ \citenamefont
  {Clark}(2005)}]{Dutton2005}%
  \BibitemOpen
  \bibfield  {author} {\bibinfo {author} {\bibfnamefont {Z.}~\bibnamefont
  {Dutton}}\ and\ \bibinfo {author} {\bibfnamefont {C.~W.}\ \bibnamefont
  {Clark}},\ }\href {https://doi.org/10.1103/PhysRevA.71.063618} {\bibfield
  {journal} {\bibinfo  {journal} {Phys. Rev. A}\ }\textbf {\bibinfo {volume}
  {71}},\ \bibinfo {pages} {063618} (\bibinfo {year} {2005})}\BibitemShut
  {NoStop}%
\bibitem [{\citenamefont {Bakkali-Hassani}\ \emph {et~al.}(2023)\citenamefont
  {Bakkali-Hassani}, \citenamefont {Maury}, \citenamefont {Stringari},
  \citenamefont {Nascimbene}, \citenamefont {Dalibard},\ and\ \citenamefont
  {Beugnon}}]{Bakkali-Hassani_2023}%
  \BibitemOpen
  \bibfield  {author} {\bibinfo {author} {\bibfnamefont {B.}~\bibnamefont
  {Bakkali-Hassani}}, \bibinfo {author} {\bibfnamefont {C.}~\bibnamefont
  {Maury}}, \bibinfo {author} {\bibfnamefont {S.}~\bibnamefont {Stringari}},
  \bibinfo {author} {\bibfnamefont {S.}~\bibnamefont {Nascimbene}}, \bibinfo
  {author} {\bibfnamefont {J.}~\bibnamefont {Dalibard}},\ and\ \bibinfo
  {author} {\bibfnamefont {J.}~\bibnamefont {Beugnon}},\ }\href
  {https://doi.org/10.1088/1367-2630/acaee3} {\bibfield  {journal} {\bibinfo
  {journal} {New Journal of Physics}\ }\textbf {\bibinfo {volume} {25}},\
  \bibinfo {pages} {013007} (\bibinfo {year} {2023})}\BibitemShut {NoStop}%
\bibitem [{\citenamefont {Romero-Ros}\ \emph {et~al.}(2022)\citenamefont
  {Romero-Ros}, \citenamefont {Katsimiga}, \citenamefont {Mistakidis},
  \citenamefont {Prinari}, \citenamefont {Biondini}, \citenamefont
  {Schmelcher},\ and\ \citenamefont {Kevrekidis}}]{Romero_theory}%
  \BibitemOpen
  \bibfield  {author} {\bibinfo {author} {\bibfnamefont {A.}~\bibnamefont
  {Romero-Ros}}, \bibinfo {author} {\bibfnamefont {G.~C.}\ \bibnamefont
  {Katsimiga}}, \bibinfo {author} {\bibfnamefont {S.~I.}\ \bibnamefont
  {Mistakidis}}, \bibinfo {author} {\bibfnamefont {B.}~\bibnamefont {Prinari}},
  \bibinfo {author} {\bibfnamefont {G.}~\bibnamefont {Biondini}}, \bibinfo
  {author} {\bibfnamefont {P.}~\bibnamefont {Schmelcher}},\ and\ \bibinfo
  {author} {\bibfnamefont {P.~G.}\ \bibnamefont {Kevrekidis}},\ }\href
  {https://doi.org/10.1103/PhysRevA.105.053306} {\bibfield  {journal} {\bibinfo
   {journal} {Phys. Rev. A}\ }\textbf {\bibinfo {volume} {105}},\ \bibinfo
  {pages} {053306} (\bibinfo {year} {2022})}\BibitemShut {NoStop}%
\bibitem [{\citenamefont {Hoefer}\ \emph {et~al.}(2006)\citenamefont {Hoefer},
  \citenamefont {Ablowitz}, \citenamefont {Coddington}, \citenamefont
  {Cornell}, \citenamefont {Engels},\ and\ \citenamefont
  {Schweikhard}}]{DSWPRAhoefer}%
  \BibitemOpen
  \bibfield  {author} {\bibinfo {author} {\bibfnamefont {M.~A.}\ \bibnamefont
  {Hoefer}}, \bibinfo {author} {\bibfnamefont {M.~J.}\ \bibnamefont
  {Ablowitz}}, \bibinfo {author} {\bibfnamefont {I.}~\bibnamefont
  {Coddington}}, \bibinfo {author} {\bibfnamefont {E.~A.}\ \bibnamefont
  {Cornell}}, \bibinfo {author} {\bibfnamefont {P.}~\bibnamefont {Engels}},\
  and\ \bibinfo {author} {\bibfnamefont {V.}~\bibnamefont {Schweikhard}},\
  }\href {https://doi.org/10.1103/PhysRevA.74.023623} {\bibfield  {journal}
  {\bibinfo  {journal} {Phys. Rev. A}\ }\textbf {\bibinfo {volume} {74}},\
  \bibinfo {pages} {023623} (\bibinfo {year} {2006})}\BibitemShut {NoStop}%
\bibitem [{\citenamefont {Mossman}\ \emph {et~al.}(2018)\citenamefont
  {Mossman}, \citenamefont {Hoefer}, \citenamefont {Julien}, \citenamefont
  {Kevrekidis},\ and\ \citenamefont {Engels}}]{DSWPRAhoefer2}%
  \BibitemOpen
  \bibfield  {author} {\bibinfo {author} {\bibfnamefont {M.~E.}\ \bibnamefont
  {Mossman}}, \bibinfo {author} {\bibfnamefont {M.~A.}\ \bibnamefont {Hoefer}},
  \bibinfo {author} {\bibfnamefont {K.}~\bibnamefont {Julien}}, \bibinfo
  {author} {\bibfnamefont {P.~G.}\ \bibnamefont {Kevrekidis}},\ and\ \bibinfo
  {author} {\bibfnamefont {P.}~\bibnamefont {Engels}},\ }\href
  {https://doi.org/10.1038/s41467-018-07147-4} {\bibfield  {journal} {\bibinfo
  {journal} {Nature Communications}\ }\textbf {\bibinfo {volume} {9}},\
  \bibinfo {pages} {4665} (\bibinfo {year} {2018})}\BibitemShut {NoStop}%
\bibitem [{sup()}]{supplement}%
  \BibitemOpen
  \href@noop {} {}\bibinfo {howpublished} {See Supplemental Material at [url] for additional details and complementary results.}\BibitemShut {Stop}%
\bibitem [{\citenamefont {Salasnich}\ \emph {et~al.}(2002)\citenamefont
  {Salasnich}, \citenamefont {Parola},\ and\ \citenamefont
  {Reatto}}]{Salasnich2002}%
  \BibitemOpen
  \bibfield  {author} {\bibinfo {author} {\bibfnamefont {L.}~\bibnamefont
  {Salasnich}}, \bibinfo {author} {\bibfnamefont {A.}~\bibnamefont {Parola}},\
  and\ \bibinfo {author} {\bibfnamefont {L.}~\bibnamefont {Reatto}},\ }\href
  {https://doi.org/10.1103/physreva.65.043614} {\bibfield  {journal} {\bibinfo
  {journal} {Physical Review A}\ }\textbf {\bibinfo {volume} {65}},\ \bibinfo
  {pages} {043614} (\bibinfo {year} {2002})}\BibitemShut {NoStop}%
\bibitem [{\citenamefont {Nicolin}\ and\ \citenamefont
  {Raportaru}(2010)}]{NICOLIN20104663}%
  \BibitemOpen
  \bibfield  {author} {\bibinfo {author} {\bibfnamefont {A.~I.}\ \bibnamefont
  {Nicolin}}\ and\ \bibinfo {author} {\bibfnamefont {M.~C.}\ \bibnamefont
  {Raportaru}},\ }\href
  {https://doi.org/https://doi.org/10.1016/j.physa.2010.06.029} {\bibfield
  {journal} {\bibinfo  {journal} {Physica A: Statistical Mechanics and its
  Applications}\ }\textbf {\bibinfo {volume} {389}},\ \bibinfo {pages} {4663}
  (\bibinfo {year} {2010})}\BibitemShut {NoStop}%
\bibitem [{\citenamefont {Pitaevskii}\ and\ \citenamefont
  {Stringari}(2003)}]{Pitaevskii2003}%
  \BibitemOpen
  \bibfield  {author} {\bibinfo {author} {\bibfnamefont {L.~P.}\ \bibnamefont
  {Pitaevskii}}\ and\ \bibinfo {author} {\bibfnamefont {S.}~\bibnamefont
  {Stringari}},\ }\href@noop {} {\emph {\bibinfo {title} {Bose-{{Einstein}}
  Condensation}}},\ \bibinfo {series} {Oxford Science Publications}\ No.\
  \bibinfo {number} {116}\ (\bibinfo  {publisher} {{Clarendon Press}},\
  \bibinfo {address} {{Oxford}},\ \bibinfo {year} {2003})\BibitemShut {NoStop}%
\bibitem [{\citenamefont {Tommasini}\ \emph {et~al.}(2003)\citenamefont
  {Tommasini}, \citenamefont {de~Passos}, \citenamefont {de~Toledo~Piza},
  \citenamefont {Hussein},\ and\ \citenamefont {Timmermans}}]{Tommasini}%
  \BibitemOpen
  \bibfield  {author} {\bibinfo {author} {\bibfnamefont {P.}~\bibnamefont
  {Tommasini}}, \bibinfo {author} {\bibfnamefont {E.~J.~V.}\ \bibnamefont
  {de~Passos}}, \bibinfo {author} {\bibfnamefont {A.~F.~R.}\ \bibnamefont
  {de~Toledo~Piza}}, \bibinfo {author} {\bibfnamefont {M.~S.}\ \bibnamefont
  {Hussein}},\ and\ \bibinfo {author} {\bibfnamefont {E.}~\bibnamefont
  {Timmermans}},\ }\href {https://doi.org/10.1103/PhysRevA.67.023606}
  {\bibfield  {journal} {\bibinfo  {journal} {Phys. Rev. A}\ }\textbf {\bibinfo
  {volume} {67}},\ \bibinfo {pages} {023606} (\bibinfo {year}
  {2003})}\BibitemShut {NoStop}%
\bibitem [{\citenamefont {Kasamatsu}\ and\ \citenamefont
  {Tsubota}(2006)}]{Kasamatsu_MI1}%
  \BibitemOpen
  \bibfield  {author} {\bibinfo {author} {\bibfnamefont {K.}~\bibnamefont
  {Kasamatsu}}\ and\ \bibinfo {author} {\bibfnamefont {M.}~\bibnamefont
  {Tsubota}},\ }\href {https://doi.org/10.1103/PhysRevA.74.013617} {\bibfield
  {journal} {\bibinfo  {journal} {Phys. Rev. A}\ }\textbf {\bibinfo {volume}
  {74}},\ \bibinfo {pages} {013617} (\bibinfo {year} {2006})}\BibitemShut
  {NoStop}%
\bibitem [{\citenamefont {Kasamatsu}\ and\ \citenamefont
  {Tsubota}(2004)}]{Kasamatsu_MI2}%
  \BibitemOpen
  \bibfield  {author} {\bibinfo {author} {\bibfnamefont {K.}~\bibnamefont
  {Kasamatsu}}\ and\ \bibinfo {author} {\bibfnamefont {M.}~\bibnamefont
  {Tsubota}},\ }\href {https://doi.org/10.1103/PhysRevLett.93.100402}
  {\bibfield  {journal} {\bibinfo  {journal} {Phys. Rev. Lett.}\ }\textbf
  {\bibinfo {volume} {93}},\ \bibinfo {pages} {100402} (\bibinfo {year}
  {2004})}\BibitemShut {NoStop}%
\bibitem [{\citenamefont {Olver}\ \emph {et~al.}(2010)\citenamefont {Olver},
  \citenamefont {Lozier}, \citenamefont {Boisvert},\ and\ \citenamefont
  {Clark}}]{NIST}%
  \BibitemOpen
  \bibfield  {author} {\bibinfo {author} {\bibfnamefont {F.~W.}\ \bibnamefont
  {Olver}}, \bibinfo {author} {\bibfnamefont {D.~W.}\ \bibnamefont {Lozier}},
  \bibinfo {author} {\bibfnamefont {D.~W.}\ \bibnamefont {Boisvert}},\ and\
  \bibinfo {author} {\bibfnamefont {C.~W.}\ \bibnamefont {Clark}},\ }\href@noop
  {} {\emph {\bibinfo {title} {{NIST} Handbook of Mathematical Functions}}}\
  (\bibinfo {year} {2010})\BibitemShut {NoStop}%
\bibitem [{\citenamefont {Biondini}(2018)}]{PRE2018v98p052220}%
  \BibitemOpen
  \bibfield  {author} {\bibinfo {author} {\bibfnamefont {G.}~\bibnamefont
  {Biondini}},\ }\href
  {http://www.math.buffalo.edu/~biondini/papers/pre2018v98p052220.pdf}
  {\bibfield  {journal} {\bibinfo  {journal} {Phys. Rev. E}\ }\textbf {\bibinfo
  {volume} {98}},\ \bibinfo {pages} {052220} (\bibinfo {year}
  {2018})}\BibitemShut {NoStop}%
\bibitem [{\citenamefont {Lannig}\ \emph {et~al.}(2020)\citenamefont {Lannig},
  \citenamefont {Schmied}, \citenamefont {Pr\"ufer}, \citenamefont {Kunkel},
  \citenamefont {Strohmaier}, \citenamefont {Strobel}, \citenamefont
  {Gasenzer}, \citenamefont {Kevrekidis},\ and\ \citenamefont
  {Oberthaler}}]{Lannig}%
  \BibitemOpen
  \bibfield  {author} {\bibinfo {author} {\bibfnamefont {S.}~\bibnamefont
  {Lannig}}, \bibinfo {author} {\bibfnamefont {C.-M.}\ \bibnamefont {Schmied}},
  \bibinfo {author} {\bibfnamefont {M.}~\bibnamefont {Pr\"ufer}}, \bibinfo
  {author} {\bibfnamefont {P.}~\bibnamefont {Kunkel}}, \bibinfo {author}
  {\bibfnamefont {R.}~\bibnamefont {Strohmaier}}, \bibinfo {author}
  {\bibfnamefont {H.}~\bibnamefont {Strobel}}, \bibinfo {author} {\bibfnamefont
  {T.}~\bibnamefont {Gasenzer}}, \bibinfo {author} {\bibfnamefont {P.~G.}\
  \bibnamefont {Kevrekidis}},\ and\ \bibinfo {author} {\bibfnamefont {M.~K.}\
  \bibnamefont {Oberthaler}},\ }\href
  {https://doi.org/10.1103/PhysRevLett.125.170401} {\bibfield  {journal}
  {\bibinfo  {journal} {Phys. Rev. Lett.}\ }\textbf {\bibinfo {volume} {125}},\
  \bibinfo {pages} {170401} (\bibinfo {year} {2020})}\BibitemShut {NoStop}%
\bibitem [{\citenamefont {El}\ \emph {et~al.}(2016)\citenamefont {El},
  \citenamefont {Khamis},\ and\ \citenamefont {Tovbis}}]{NLTY2016v29p2798}%
  \BibitemOpen
  \bibfield  {author} {\bibinfo {author} {\bibfnamefont {G.~A.}\ \bibnamefont
  {El}}, \bibinfo {author} {\bibfnamefont {E.~G.}\ \bibnamefont {Khamis}},\
  and\ \bibinfo {author} {\bibfnamefont {A.}~\bibnamefont {Tovbis}},\
  }\href@noop {} {\bibfield  {journal} {\bibinfo  {journal} {Nonlinearity}\
  }\textbf {\bibinfo {volume} {29}},\ \bibinfo {pages} {2798} (\bibinfo {year}
  {2016})}\BibitemShut {NoStop}%
\bibitem [{\citenamefont {Audo}\ \emph {et~al.}(2018)\citenamefont {Audo},
  \citenamefont {Kibler}, \citenamefont {Fatome},\ and\ \citenamefont
  {Finot}}]{Audo:18}%
  \BibitemOpen
  \bibfield  {author} {\bibinfo {author} {\bibfnamefont {F.}~\bibnamefont
  {Audo}}, \bibinfo {author} {\bibfnamefont {B.}~\bibnamefont {Kibler}},
  \bibinfo {author} {\bibfnamefont {J.}~\bibnamefont {Fatome}},\ and\ \bibinfo
  {author} {\bibfnamefont {C.}~\bibnamefont {Finot}},\ }\href
  {https://doi.org/10.1364/OL.43.002864} {\bibfield  {journal} {\bibinfo
  {journal} {Opt. Lett.}\ }\textbf {\bibinfo {volume} {43}},\ \bibinfo {pages}
  {2864} (\bibinfo {year} {2018})}\BibitemShut {NoStop}%
\bibitem [{\citenamefont {Peregrine}(1983)}]{peregrine1983water}%
  \BibitemOpen
  \bibfield  {author} {\bibinfo {author} {\bibfnamefont {D.~H.}\ \bibnamefont
  {Peregrine}},\ }\href@noop {} {\bibfield  {journal} {\bibinfo  {journal} {The
  ANZIAM Journal}\ }\textbf {\bibinfo {volume} {25}},\ \bibinfo {pages} {16}
  (\bibinfo {year} {1983})}\BibitemShut {NoStop}%
\bibitem [{\citenamefont {Bertola}\ and\ \citenamefont
  {Tovbis}(2013)}]{bertola}%
  \BibitemOpen
  \bibfield  {author} {\bibinfo {author} {\bibfnamefont {M.}~\bibnamefont
  {Bertola}}\ and\ \bibinfo {author} {\bibfnamefont {A.}~\bibnamefont
  {Tovbis}},\ }\href {https://doi.org/https://doi.org/10.1002/cpa.21445}
  {\bibfield  {journal} {\bibinfo  {journal} {Commun. Pure Appl. Math.}\
  }\textbf {\bibinfo {volume} {66}},\ \bibinfo {pages} {678} (\bibinfo {year}
  {2013})}\BibitemShut {NoStop}%
\bibitem [{\citenamefont {Navon}\ \emph {et~al.}(2016)\citenamefont {Navon},
  \citenamefont {Gaunt}, \citenamefont {Smith},\ and\ \citenamefont
  {Hadzibabic}}]{navon2016emergence}%
  \BibitemOpen
  \bibfield  {author} {\bibinfo {author} {\bibfnamefont {N.}~\bibnamefont
  {Navon}}, \bibinfo {author} {\bibfnamefont {A.~L.}\ \bibnamefont {Gaunt}},
  \bibinfo {author} {\bibfnamefont {R.~P.}\ \bibnamefont {Smith}},\ and\
  \bibinfo {author} {\bibfnamefont {Z.}~\bibnamefont {Hadzibabic}},\
  }\href@noop {} {\bibfield  {journal} {\bibinfo  {journal} {Nature}\ }\textbf
  {\bibinfo {volume} {539}},\ \bibinfo {pages} {72} (\bibinfo {year}
  {2016})}\BibitemShut {NoStop}%
\bibitem [{\citenamefont {Navon}\ \emph {et~al.}(2021)\citenamefont {Navon},
  \citenamefont {Smith},\ and\ \citenamefont {Hadzibabic}}]{navon2021quantum}%
  \BibitemOpen
  \bibfield  {author} {\bibinfo {author} {\bibfnamefont {N.}~\bibnamefont
  {Navon}}, \bibinfo {author} {\bibfnamefont {R.~P.}\ \bibnamefont {Smith}},\
  and\ \bibinfo {author} {\bibfnamefont {Z.}~\bibnamefont {Hadzibabic}},\
  }\href@noop {} {\bibfield  {journal} {\bibinfo  {journal} {Nature Phys.}\
  }\textbf {\bibinfo {volume} {17}},\ \bibinfo {pages} {1334} (\bibinfo {year}
  {2021})}\BibitemShut {NoStop}%
\end{thebibliography}
\end{document}